\documentclass[twocolumn]{aastex631}

\newcommand{\hi}{H {\sc i}}
\newcommand{\hii}{H {\sc ii}}
\newcommand{\hei}{He {\sc i}}
\newcommand{\heii}{He {\sc ii}}
\newcommand{\heiii}{He {\sc iii}}

\usepackage{physics}  
\usepackage{orcidlink}

\begin{document}

\title{A Hierarchical Shock Model of Ultra-High-Energy Cosmic Rays}

\author[0000-0001-7763-4405]{Paul Simeon}
\affiliation{Physics Department, Stanford University,
382 Via Pueblo Mall, 
Stanford, CA 94305, USA}
\affiliation{Kavli Institute for Particle Astrophysics and Cosmology, Stanford University, Stanford, CA 94305, USA}
\correspondingauthor{Paul Simeon}
\email{paulsimeon@gmail.com}

\author[0000-0001-9011-0737]{N. Globus}
\affiliation{Kavli Institute for Particle Astrophysics and Cosmology, Stanford University, Stanford, CA 94305, USA}
\affiliation{Instituto de Astronomía, Universidad Nacional Autónoma de México Campus Ensenada, Ensenada, BC 22800, México}
\affiliation{Astrophysical Big Bang Laboratory, RIKEN, Wako, Saitama, Japan}

\author[0000-0002-8638-1697]{Kirk S. S. Barrow}
\affiliation{Astronomy Department, College of Liberal Arts and Sciences,  University of Illinois at Urbana-Champaign,  Urbana, IL 61801, USA}

\author[0000-0002-1854-5506]{Roger Blandford}
\affiliation{Kavli Institute for Particle Astrophysics and Cosmology, Stanford University, Stanford, CA 94305, USA}
\affiliation{SLAC National Accelerator Laboratory,
2575 Sand Hill Road, Menlo Park, CA 94025, USA}

\begin{abstract} 

We propose that a hierarchical shock model---including supernova remnant shocks, galactic wind termination shocks, and accretion shocks around cosmic filaments and galaxy clusters---can naturally explain the cosmic ray spectrum from $\sim$1\,GeV up to $\sim$\,200\,EeV. While this framework applies to the entire cosmic ray spectrum, in this work, we focus on its implications for ultra-high-energy cosmic rays (UHECRs). We perform a hydrodynamic cosmological simulation to investigate the power processed at shocks around clusters and filaments. The downstream flux from nearby shocks around the local filament accounts for the softer, lower-energy extragalactic component around the ankle, and the upstream escaping flux from nearby clusters accounts for the transition to a hard spectral component at the highest energies. This interpretation is in agreement with UHECR observations. We suggest that a combination of early-Universe galactic outflows, cosmic ray streaming instabilities, and a small-scale turbulent dynamo can increase magnetic fields enough to attain the required rigidities. Our simulation suggests that the available volume-averaged power density of accretion shocks exceeds the required UHECR luminosity density by three orders of magnitude. We show that microgauss magnetic fields at these shocks could explain both the origin of UHECRs and potentially contribute to the diffuse radio synchrotron background below 10\,GHz. The shock-accelerated electrons produce a hard radio background without overproducing diffuse inverse Compton emission. These results motivate further observational tests with upcoming facilities to help distinguish accretion shocks from other UHECR sources.
 
\end{abstract}

\keywords{Cosmic ray sources (328), Ultra-high-energy cosmic radiation (1733), Cosmic web (330), Shocks (2086)}

\section{Introduction} \label{sec:intro}

The origin of ultra-high-energy cosmic rays (UHECRs)---cosmic rays with energies above $10^{18}$\,eV, or 1\,EeV---remains unknown despite decades of observational efforts. Relativistic jets from active galactic nuclei (AGNs), gamma-ray bursts, tidal disruption events, ultra-fast outflows and neutron star mergers are popular source candidates because their extreme energy budgets, high magnetic fields, and relativistic outflows allow efficient acceleration up to the highest rigidities ($\sim$10 EV\footnote{We define rigidity as $R \equiv pc/Ze \simeq {E}/{Ze}$, which has units of volts and determines how particles interact with magnetic fields. Cosmic ray acceleration and diffusion depend on rigidity rather than energy.}; \citealt{2025ARA&A..63..339G}). However, with our current understanding of source candidates and their environments, no single existing model can conclusively explain all UHECR observations \citep[e.g.,][]{2023EPJWC.28304001G}. Modeling the full cosmic ray spectrum, particularly the transition from Galactic to extragalactic components, remains a major challenge.

Diffusive shock acceleration (DSA) is one of the best mechanisms to accelerate particles, and we suggest it is natural to look for increasingly large nonrelativistic, collisionless astrophysical shocks to accelerate cosmic rays in a multi-stage process to create a relatively smooth spectrum over the full energy range. The largest, most suitable shocks for accelerating nuclei to energies above 100\,EeV (extreme energy cosmic rays, EECRs) are the accretion shocks---also known as virial shocks or structure-formation shocks---formed as cold gas accretes onto the large-scale structure of galaxy clusters and filaments. These shocks, which are megaparsecs in size, have existed continuously for billions of years, forming an invisible boundary around virtually all large-scale structure.

The many models for the acceleration of cosmic rays at large-scale accretion shocks vary in the cosmic ray energy range they address, in UHECR composition, whether they focus on shocks around galaxy clusters or galaxy filaments, and in other aspects \citep{1995ApJ...454...60N,1996ApJ...456..422K,2007astro.ph..1167I,2008ApJ...689L.105M,2009BRASP..73..552P,2014PhDT.......656S,2019JPhCS1181a2033Z}. In all these models, the maximum energy $E_{\rm max} \propto Z\,B\,u_{1}$ requires strong magnetic turbulence ($B_{1} \sim 1\,\mu{\rm G}$) upstream of the shock front with a sufficiently high shock speed (typically $u_{1} \sim 1000\,{\rm km\,s}^{-1}$) to lower the acceleration timescale below any energy-loss timescale. Heavy nuclei with atomic number $Z$ increase the maximum energy, consistent with the heavier composition reported by the Pierre Auger Collaboration (Auger) \citep{2024PhRvD.109j2001A}. The main requirement for accretion shocks is the presence of sufficiently strong magnetic turbulence ahead of the shock fronts. We contend that the magnetic turbulence builds up over time through the combination of galactic outflows in the early universe, cosmic-ray-induced kinetic instabilities, and a pressure-driven turbulent dynamo. 

We present a hierarchical shock acceleration model that combines several ideas in a multiscale hierarchy of strong, collisionless shocks, where the spectrum of cosmic rays from one class of shocks depends on the escape of lower-energy cosmic rays from shocks lower in the hierarchy. Cosmic rays first escape from supernova remnant shocks within galaxies and then propagate outward into galactic wind termination shocks. From there, they reach filamentary accretion shocks that surround most galaxies and, in rare cases, continue to cluster accretion shocks, where EECRs can reach up to $\sim$200\,EeV. At cluster accretion shocks, the composition of the accelerated cosmic rays remains mixed, reflecting their galactic origins. However, iron nuclei dominate EECRs. Nonrelativistic DSA can then account for the majority of the cosmic ray spectrum.

Observations of radio, X-ray, and gamma-ray emission from nonthermal electrons accelerated in these shocks support this model. Nonthermal electrons with a power-law index $p\sim 2.2$ downstream of the shock emit synchrotron radiation with a spectral index $\alpha = (p-1)/2 \sim 0.6$, matching the observed isotropic radio synchrotron background between 22\,MHz and 10\,GHz \citep{2011ApJ...734....5F,2011ApJ...734....6S}.
Current models of known radio populations do not fully explain this background, suggesting that large-scale structure shocks could provide an additional contribution. If diffuse electrons produce this emission without overproducing X-ray and gamma-ray background, the electrons must be in magnetic fields of at least 1\,$\mu$G \citep{2010MNRAS.409.1172S}, which is also required for these shocks to produce the highest-energy cosmic rays. The same unexpectedly large magnetic fields---which we estimate to be plausible---can connect two puzzles at opposite ends of the multimessenger energy spectrum. If UHECRs originate from accretion shocks, they should produce a distinct spectrum of cosmogenic neutrinos compared to AGNs or gamma-ray bursts, reflecting differences in the redshift evolution of the source power. Such neutrino emission from cluster accretion shocks has been modeled in detail by \citet{2016ApJ...828...37F}, who predict a subdominant but detectable contribution to the diffuse neutrino background. This distinction could be tested with upcoming neutrino observatories such as GRAND.

In Section\,\ref{sec:sims} we describe a hydrodynamic simulation of accretion shocks around large halos, which we use to estimate the cosmic ray luminosity of individual clusters and the cosmological average. Although this simulation does not include magnetic fields or cosmic ray feedback, it effectively captures the energetics of the shocks. In Section\,\ref{sec:magnetic-fields}, we propose mechanisms for magnetic field amplification and the maximum cosmic ray energy achievable at accretion shocks. In Section\,\ref{sec:hierarchcal-model}, we present our hierarchical shock model and its implications for the cosmic ray spectrum and composition. Section\,\ref{sec:observations} explores the observational constraints on this model and examines the potential connection to the radio synchrotron background. We discuss the broader implications of this model in Section\,\ref{sec:discussion} and conclude in Section\,\ref{sec:conclusion}.

\section{Hydrodynamic simulation of accretion shocks}
\label{sec:sims}

\subsection{Methods}

We performed a radiation hydrodynamic adaptive-mesh refinement (AMR) {\sc Enzo} \citep{2014ApJS..211...19B} simulation to predict accretion shock dynamics around large-scale structures with higher resolution and more realistic radiation backgrounds than previous work. We initialized a box with a side length of 256 co-moving Mpc\,$h^{-1}$ with a 256$^3$ root grid of dark matter particles (mass resolution of $7.17 \times 10^{10}\ M\rm{_\odot}$/particle) and baryons at $z = 100$ using the code {\sc Music} \citep{2011MNRAS.415.2101H}. We assumed a spatially flat cosmology with parameters: $\Omega_{\rm M} = 0.3065$, $\Omega_{\Lambda} = 0.6935 $, $\Omega_{\rm b} = 0.0483$, $h = 0.679$, $\sigma_8 = 0.8154$, and $n = 0.9681$ \citep{2016A&A...594A..13P}. The simulation evolves from these initial conditions to $z=0$ with outputs every $\sim$10\,Myr during the final billion years.  

We identified temperature discontinuities across simulation cells with an ab initio shock-finding algorithm \citep{2008ApJ...689.1063S} and recorded preshock temperatures, densities, and Mach numbers. We employed a \citet{2012ApJ...746..125H} background to radiatively drive a 9-species (\hi, \hii, \hei, \heii, \heiii, e$^-$, H$_2$, H$_2^+$, H$^-$) non-equilibrium chemistry throughout the intergalactic medium. This radiation background eliminates the need for a preset shock-finding temperature floor used in previous studies and facilitates a more reliable determination of shock speeds. Additionally, baryons can refine up to nine AMR levels, achieving a maximum gas spatial resolution of $\sim$1.95 co-moving kpc $h^{-1}$, sufficient to resolve galactic disk structures within clusters. The simulation can resolve the dark matter gravitational potential of a $10^{14}\, \rm{M_\odot}$ galaxy cluster with more than 1400 particles. We constructed halo merger trees using the codes {\sc Rockstar} \citep{2013ApJ...762..109B} and {\sc Consistent Trees} \citep{2013ApJ...763...18B}, which use simulation dark matter particles to determine halo-like associations in phase space. 

In this analysis, we selected shocks with Mach numbers greater than 5 to keep only strong shocks that are efficient particle accelerators, and we excluded shocks deep within clusters and galaxies by selecting only shocks with preshock overdensity $\delta = \rho_{\rm b} / \langle \rho_{\rm b} \rangle < 10^{3}$, where $\rho_{\rm b}$ is the baryon density. These criteria reveal a web of strong shocks in low-density regions surrounding the large-scale structure of clusters and filaments.

\begin{figure*}[htb!]
    \centering
    \includegraphics[width=\linewidth]{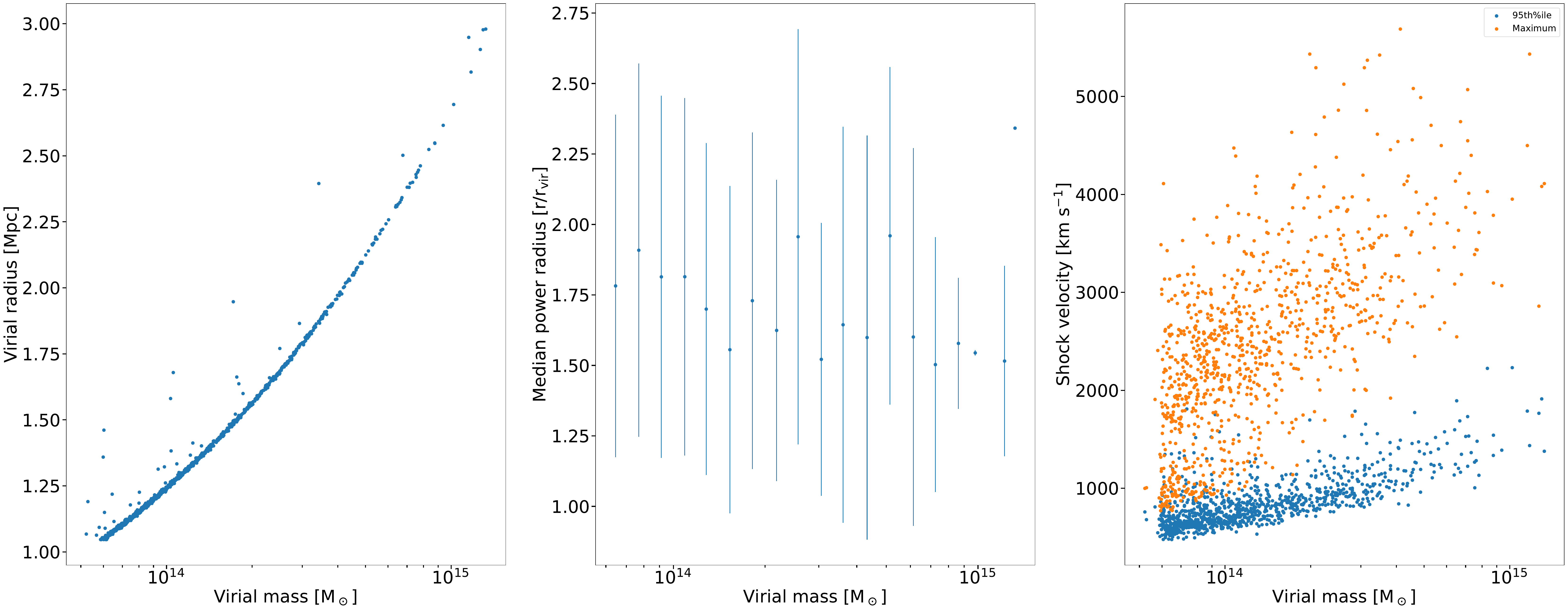}
    \caption{Left: virial radius for each halo in the simulation. Center: the radius for each halo that contains half the total power being dissipated by shocks that pass our cuts. Error bars represent one standard deviation above and below the mean. Right: the maximum (orange) and the 95th percentile (blue) upstream speed, $u_1$, taken from the histogram of shock velocities for each halo.}
    \label{fig:shock-properties}
\end{figure*}

\subsection{Simulation results}
\label{sec:sim_results}

Our simulation results align with previous studies of large-scale shocks (e.g., \citealt{2008ApJ...689.1063S,2009MNRAS.395.1333V}). However, our focus is on whether these shocks can efficiently accelerate cosmic rays, rather than their role in structure formation. Our simulation reveals a large cosmic web of strong shocks at $z = 1$ with similar size and power as those at $z = 0$, and the accretion shocks around individual clusters remained stable throughout this period. If accretion shocks accelerate cosmic rays, then those cosmic rays could have been gaining energy for over 8\,Gyr. This long acceleration time is a distinct advantage of accretion shocks over other cosmic ray sources.

The source of energy for accelerating cosmic rays is the kinetic energy processed through the shocks per unit time. The volume-averaged power density of high-Mach, low-density shocks identified in our simulation is $\mathcal{P}_{\rm LSSS} \approx 1.0 \times 10^{40}$~erg~s$^{-1}$~Mpc$^{-3}$ ($3.4 \times 10^{-35}$~W~m$^{-3}$), exceeding the required UHECR luminosity density above 1\,EeV by three orders of magnitude: $\mathcal{L}_{\rm UHECR} \sim 3 \times 10^{36}\,{\rm erg\,s}^{-1}\,{\rm Mpc}^{-3}$ ($10^{-38}\,{\rm W\,m}^{-3}$). 

Throughout this paper, we measure galaxy clusters by their virial mass ($M_{\rm vir} \equiv M_{200}$), defined as the mass enclosed within $r_{200}$, which is the radius where the enclosed density is 200 times the critical density of the universe at that redshift. For each halo in the simulation, the virial radius lies on or above the virialization trend $r_{\rm vir}\sim1.25\,M_{14}^{1/3}$~Mpc, where $M_{14} \equiv M_{200}/(10^{14} M_{\odot})$ (Figure~\ref{fig:shock-properties}, left panel). Cluster accretion shocks are not perfectly spherical. However, the radius at which half of the shock power is enclosed is roughly a factor 1.5--2 times $r_{\rm vir}$, as seen in the center panel of Figure~\ref{fig:shock-properties}. This radius is a reasonable approximation for the shock radius for most halos where the majority of the power is dissipated sharply at a single quasi-spherical accretion shock. In this paper, we take the shock radius to be $r_{\rm sh} \approx 2 r_{\rm vir}$ for convenience. Some cylindrical filaments have a radius of 1--2\,Mpc, but the shapes are irregular with widening or narrowing filaments and concave shock surfaces as the filaments merge with clusters.

\begin{figure*}[htb!]
    \centering
    \includegraphics[width=\linewidth]{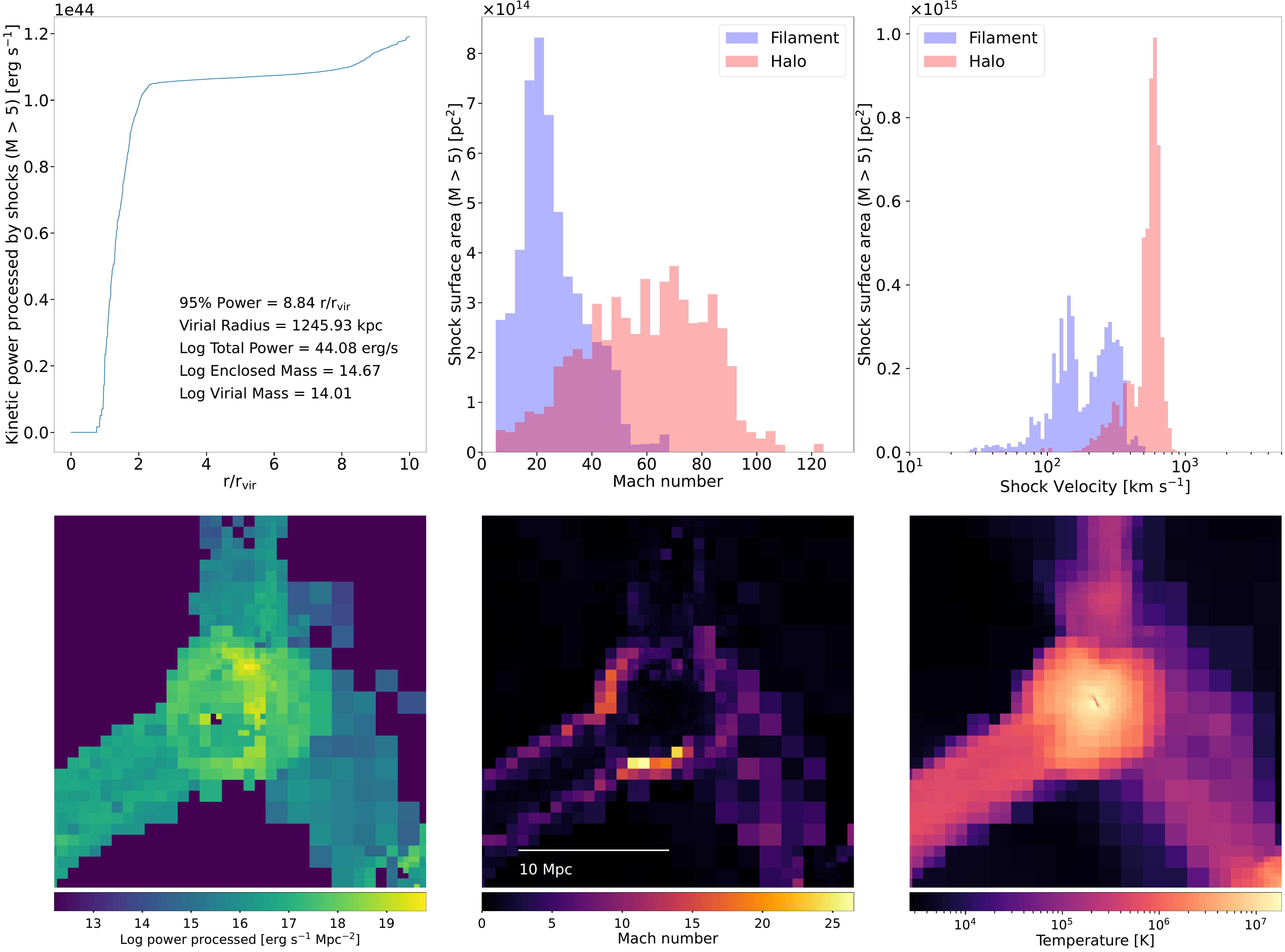}
    \caption{Virgo analog halo, selected for its roughly spherical accretion shock with attached filaments and a virial mass similar to the Virgo Cluster. In this analysis, the virial radius was defined as $r_{200}$ and the virial mass as $M_{200}$. All shocks were selected with $M >5$ and preshock overdensity less than $10^3$. Top left: cumulative kinetic energy processed by shocks as a function of radius. Top center: cumulative surface area weighted by Mach number. Top right: histogram of shock surface area vs. shock speed ($u_1$). Shocks within $3 r_{\rm vir}$ of any halo larger than 1\% of the central halo are labeled ``Halo.'' All other shocks were labeled as ``Filament." Bottom left: projection of logarithmic energy processed by shocks. Bottom center: projection of Mach number of selected shocks. Mach numbers appear lower than their true values due to volume-averaged projection effects. Bottom right: projection of gas temperature.}
    \label{fig:546Virgo}
\end{figure*}

\begin{figure*}[htb!]
    \centering
    \includegraphics[width=\linewidth]{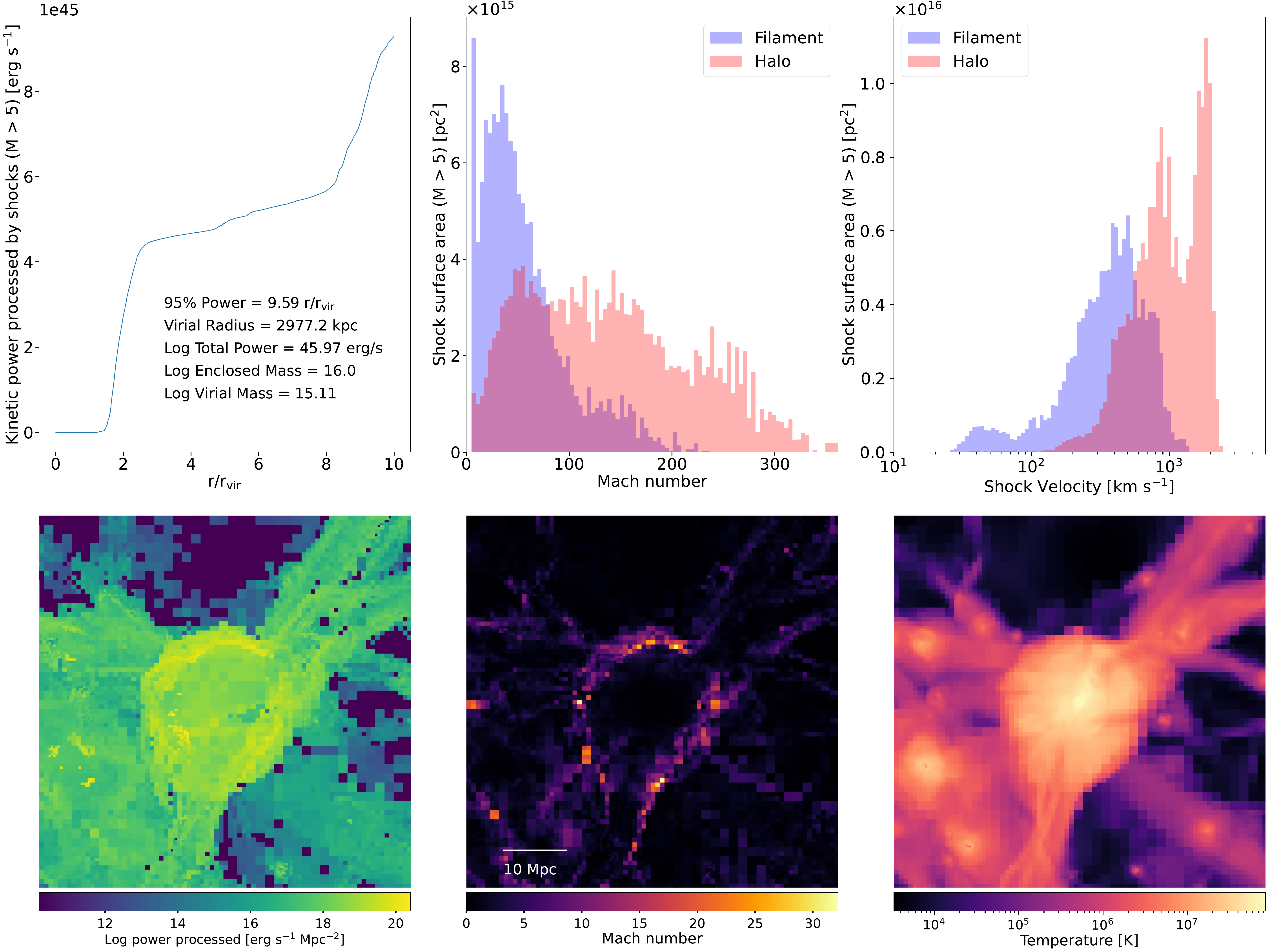}
    \caption{Coma analog halo, the most massive halo in the simulation box, with a virial mass of $10^{15.11} M_{\odot}$ comparable to the Coma Cluster.}
    \label{fig:sim_1Coma}
\end{figure*}

The shocks in our simulation show two distinct velocity distributions corresponding to the two primary geometries formed by gravitationally accreting matter: clusters and filaments. The quasi-spherical shocks around clusters form nodes in the cosmic web, where matter is most concentrated, making them the strongest type of accretion shock. The quasi-cylindrical shocks around filaments are weaker than cluster shocks because matter is extended in one dimension, reducing the depth of the gravitational well. Large sheets can also form during gravitational collapse, but we did not investigate them in this work. See the top right panels of Figures~\ref{fig:546Virgo} and~\ref{fig:sim_1Coma}, which show the full distributions of shock velocities for two representative halos.

The upstream gas speed, $u_1$, relative to the shock front is typically around 1000\,km\,s$^{-1}$ for clusters with $M_{14} \geq 1$, but small regions of the shock fronts can reach up to 5000\,km\,s$^{-1}$, as seen in the right panel of Figure~\ref{fig:shock-properties}. We plotted the maximum (orange) and the 95th percentile (blue) $u_1$ taken from the histogram of shock velocities for each halo. The 95th percentile speed better represents typical cluster shocks since most of a halo's shock surface area consists of weaker filament shocks\footnote{The median or mean of the non-Gaussian velocity distribution would likely correspond to a filament shock, which is not representative of the regions capable of producing EECRs.}. The lower edge of the 95th percentile distribution of points follows $u_1 \sim 600\,M_{14}^{1/3}\,{\rm km\,s}^{-1}$. Based on the observed range of $u_1$, we adopt fiducial values of 1000\,km\,s$^{-1}$ for cluster shocks and 400\,km\,s$^{-1}$ for filament shocks.

The rare, fast regions of the shock front play a crucial role in accelerating cosmic rays above 100\,EeV. Because the acceleration time scales as $u_{1}^{-2}$, the highest-energy cosmic rays are most likely to originate from these regions, where particles can gain energy up to 20 times faster than in typical regions of the accretion shock. The more numerous, weaker shocks contribute predominantly to the lower-energy extragalactic component below the ankle and may also supply seed particles that are subsequently reaccelerated by the stronger shocks.

Now that we have characterized the overall distribution of shock properties, we examine individual clusters to illustrate how these properties vary with cluster mass and geometry. The reference cluster is the Virgo Cluster, the most important cluster for our work because of its size and proximity. \citet{2017MNRAS.469.1476S} estimate the mass and radius of the Virgo Cluster to be $M_{14}=1$ and $r_{200} = 1$\,Mpc at a distance of 16\,Mpc, though other estimates vary (e.g., \citet{2020A&A...635A.135K} estimate a higher virial mass of $M_{14} = 6$). The Coma Cluster is more massive at $M_{14} \sim 30$ \citep{2007ApJ...671.1466K}, but its distance of $\sim$100\,Mpc makes it a less prominent source of EECRs than Virgo. Additionally, Virgo's proximity makes it a more prominent source of anisotropy in the sky. The Perseus Cluster, with a mass of $M_{14} \sim 7$ and a distance of $\sim$80\,Mpc \citep{2011Sci...331.1576S}, is another large cluster within the cosmic ray horizon for UHECRs with $E\lesssim 100\,{\rm EeV}$. 

We selected a simulated halo from our simulation with a mass comparable to the Virgo Cluster and a typical morphology of three filamentary structures connecting to a spheroidal main cluster as shown in Figure~\ref{fig:546Virgo}. We refer to this halo as our Virgo analog. The top-left panel of Figure~\ref{fig:546Virgo} illustrates the cumulative kinetic energy processed as a function of radius from the center of the cluster. Although the accretion shock is thinner than the simulation resolution, deviations from spherical symmetry and the presence of filaments cause the processed kinetic energy to be spread out over a range of 1.2--2.5\,Mpc from the center of the cluster. Based on this range, we approximate the shock radius as $r_{\rm sh} \sim 2\,{\rm Mpc}$. The total power dissipated at the main accretion shock for clusters of this size is typically around $10^{44}$\,erg\,s$^{-1}$. Additionally, the total power of all of the filamentary shocks outside of the central halo is about 10\% of the power of the central halo's accretion shock. We observe a similar fraction in other clusters in the simulation and generally assume that this proportion holds for all clusters.

The top-right panel of Figure~\ref{fig:546Virgo} shows the distribution of shock surface area as a function of upstream velocity for the Virgo analog halo. The quasi-spherical accretion shocks around galaxy clusters---plotted in red and labeled ``Halo"---were selected based on their location within $\sim$3\,$r_{\rm vir}$ of any halo exceeding 1\% of the central halo's mass. These shocks have the highest speeds but the smallest surface area, which is natural for the highest concentrations of matter.

We were unable to distinguish filaments from sheets, so all shocks outside the halo shocks are labeled as ``Filament.'' Filaments have the largest surface area because most large-scale structure and shocks reside in the filamentary web and because filaments have a high surface-area-to-volume ratio. Our Virgo-sized cluster has a grouping of $u_1$ around 600--700\,km~s$^{-1}$ (Halo component) in addition to a large surface area with slower speeds around 200 km~s$^{-1}$ (Filament component). The speeds in this simulation match reasonably well to observations of the velocity dispersion of 640\,km\,s$^{-1}$, which should be lower than the shock velocity by a factor of 1.5--2 \citep{2020A&A...635A.135K}. Additionally, our simulation agrees with \citet{2017ApJ...850..207S}, who modeled the radial velocities around Virgo. The top-middle panel of Figure~\ref{fig:546Virgo} shows the distribution of shock surface area as a function of Mach number, confirming that the regions with the highest Mach number tend to be around the cluster rather than along filaments.

Figure~\ref{fig:sim_1Coma} presents the same analysis for a Coma analog halo, the most massive halo in our simulation, which also exhibits a typical spherical morphology surrounded by filaments. The integrated kinetic energy dissipated by shocks is significantly higher than in the Virgo analog, reaching a few times $10^{45}$\,erg\,s$^{-1}$. The approximate shock radius is also much larger, at $r_{\rm sh} \sim 6\,{\rm Mpc}$. This cluster, with $M_{14} \sim 10$, exhibits a high-velocity component around 1000--2000\,km\,s$^{-1}$ for the Halo shocks, which is consistent with Coma's velocity dispersion of 1000--1200\,km\,s$^{-1}$. The lower-velocity component around 400\,km\,s$^{-1}$ corresponds to the Filament shocks.

\section{Magnetic fields and maximum energy limits}
\label{sec:magnetic-fields}

\subsection{Prior estimates and limits on the intergalactic magnetic field}

The acceleration of the highest-energy cosmic rays requires relatively strong magnetic fields upstream of the accretion shocks in the cosmic web. While precise measurements of the magnetic field in these regions are lacking, our model is consistent with existing observations and upper limits on intergalactic magnetic fields.

Our model requires magnetic fields of order 1\,$\mu$G in both the upstream and downstream regions of the accretion shocks around galaxy clusters, assuming a fiducial upstream velocity of $u_1 = 1000\,{\rm km\,s}^{-1}$, and approximately 0.1\,$\mu$G upstream of filament shocks. The characteristic length scale of the upstream region is comparable to the shock radius, beyond which the cosmic rays effectively escape as the increasing upstream volume causes the cosmic ray density and magnetic turbulence to drop too low for efficient scattering back to the shock. Outside this precursor region, the field falls below 10\,nG, allowing cosmic rays to propagate more freely. This low field strength in cosmic voids is consistent with observations \citep{2024arXiv241214825N} and with studies of the propagation of UHECRs \citep{2018MNRAS.475.2519H,2019MNRAS.484.4167G}. In these voids, the dominant source of magnetic fields may be the streaming cosmic rays themselves, which induce turbulence as they escape the web.

Current observational constraints on the magnetic field in the intergalactic medium of clusters and filaments probe large scales larger than 100\,kpc, leaving open the possibility of stronger, small-scale fields. The wave modes that scatter the highest-energy cosmic rays have length scales below 10\,kpc if the magnetic field strength is 1\,$\mu$G. Rotation measure estimates of large-scale magnetic fields in cosmic filaments at $z=0$ range from 39--84\,nG \citep{2023MNRAS.518.2273C}. Estimates of the turbulent component of these fields range from 3.5 to 5.5\,nG, but only on scales of 0.4--1.0\,Mpc \citep{2022MNRAS.512..945C}. Therefore, current observations do not rule out the presence of stronger magnetic fields on smaller scales near accretion shocks.

The denser, brighter regions of the cosmic web are easier to observe, so most current measurements are well inside the accretion shock. For example, \citet{2021NatAs...5..268D} detect radio relics and halos in galaxy clusters at $z\approx 0.6$--$0.9$ that imply magnetic fields amplified above 1\,$\mu$G by cluster mergers. More recently, distributed synchrotron emission from cluster Abell\,2255 ($z = 0.08$) suggests magnetic fields between 0.1\,$\mu$G and 2\,$\mu$G at a distance of 2\,Mpc from the cluster center \citep{2022SciA....8.7623B}. Radio and gamma-ray observations of the Coma Cluster's radio halo indicate a magnetic field of $\sim$5\,$\mu$G in the core, gradually decreasing to $\sim$0.7\,$\mu$G at $r\sim 2\,{\rm Mpc}$ from the center \citep{2022ApJ...933..218B}.

\subsection{Attaining microgauss magnetic fields}
\label{subsec:magnetic-fields}

The crucial piece in this model is the high level of cosmic ray scattering that keeps the cosmic rays crossing the shock front and shortens the acceleration time. This requirement is usually expressed as the need for Bohm-like diffusion in magnetic fields of about 1\,$\mu$G on the length scales of 1\,pc to 10\,kpc, corresponding to the gyroradius of cosmic rays in the rigidity range of 1\,PV to 10\,EV. Although this length scale is too small to be constrained by observations in the outskirts of clusters and in filaments, we argue that this level of magnetic field strength could plausibly arise from the combined effects of several mechanisms, including the following three: galactic outflows in the early universe, cosmic ray streaming instabilities, and a small-scale dynamo\footnote{\citet{2006ApJ...642L...1M} considered the amplification of magnetic fields around galaxy clusters first by plasma instabilities at accretion shocks and then by dynamos. Our general approach is similar in spirit but differs in mechanism. We focus on streaming instabilities rather than the Weibel instability, and on a dynamo driven by cosmic ray pressure gradients rather than by generic turbulence or shear.} driven by cosmic ray pressure gradients acting on density inhomogeneities. However, the exact interplay of these processes remains uncertain.

The first source of magnetic fields are the continuous galactic outflows, driven by supernova explosions, which have enriched the intergalactic medium with magnetic fields since the early universe ($z\sim 3$) before the large-scale structure formed \citep{2020MNRAS.498.3125P}. Even in the absence of the flux of cosmic rays accelerated at accretion shocks, the magnetic fields upstream of those shocks could be greater than 10\,nG. \citet{2020MNRAS.498.3125P} find in their simulation a magnetic field strength of 100\,nG at the virial radius ($\sim$200\,kpc) of Milky Way-like galaxies. Although such results are often discussed in the context of the circumgalactic medium, the number density of galaxies was higher at high redshift, so their outflows magnetized large fractions of the proto-cluster IGM. This field provides the baseline level of magnetic field upstream of the accretion shocks to allow DSA and magnetic field amplification earlier and more efficiently than in a pristine IGM. The high metallicity in the outskirts of clusters implies that the gas was enriched with metals, magnetic fields, and cosmic rays over 10\,billion years ago \citep{2013Natur.502..656W}.

The second source of amplified magnetic fields is the ensemble of various plasma instabilities driven by cosmic rays upstream of their source. An anisotropic flux of cosmic rays generates a net cosmic ray current, which in turn induces a compensating return current in the background plasma. This configuration is unstable to the cosmic ray streaming instability, both the resonant \citep{1975MNRAS.172..557S} and non-resonant \citep{2004MNRAS.353..550B} branches. The resonant mode grows fastest for wavelengths resonant with the gyroradius of the cosmic rays, $\lambda_{\rm r} \sim r_{\rm g}(p)$, and so these long-wavelength fluctuations can efficiently scatter the particles. However, in the high-Mach-number shocks of accretion shocks and supernova remnant shocks, the non-resonant mode grows faster than the resonant mode \citep{2009MNRAS.392.1591A}, especially when the direction of an ordered background magnetic field is less than 45$^{\circ}$ to the cosmic ray current \citep{2014ApJ...783...91C}. The return current of the electrons, which balances the positive cosmic ray current, drives the non-resonant mode\footnote{This instability is sometimes referred to as the cosmic ray current-driven instability or the Bell instability.}. The non-resonant modes are therefore oppositely polarized and shorter in wavelength than the resonant modes, $\lambda_{\rm nr} \ll r_{\rm g}(p)$.

Initially, these short-wavelength modes do not affect the cosmic rays driving the current. However, as the instability continues to grow nonlinearly, the wavelength of the fastest-growing mode increases. Once the wavelength of the non-resonant modes becomes comparable to the gyroradii of the cosmic rays driving the instability, those modes scatter the cosmic rays, and the disrupted current ends the growth of the magnetic field \citep{2014ApJ...783...91C,2014ApJ...794...46C}. The growth timescale is short compared to cluster dynamical times. For typical upstream densities, CR fractions, and drift speeds, the growth timescale $\tau_{\rm mag}$ for both resonant and non-resonant modes is $\lesssim 10 \,{\rm Myr}$, well below the advection time across the precursor. The growth timescale for the nonresonant mode exceeds 100\,Myr when $n_{\rm CR}(>p)/n \lesssim 10^{-15}$.

Magnetic energy increases over the initial value by a factor $\left\langle \frac{B_{{\rm tot}}}{B_{0}} \right\rangle^2 \approx 3 \zeta_{\rm CR} M_{\rm A}$, where $\zeta_{\rm CR}$ is the ratio of the cosmic ray pressure at the shock to $\rho u_{1}^2$, including some power at scales longer than the gyroradius of the most energetic particles. Previous studies have shown that the non-resonant streaming instability typically amplifies the magnetic field by a factor $\lesssim 10$ in the upstream medium of the forward shocks of supernova remnants \citep{2009ApJ...694..626R,2014ApJ...794...46C}. We expect this mechanism to be effective in the upstream medium of accretion shocks, particularly when an ordered background field is present, where the low gas density and high Alfvénic Mach number yields a large value of $\zeta_{\rm CR} M_{\rm A}$. In regions where magnetic fields are initially very weak, cosmic rays can stream nearly freely, thereby driving streaming instabilities that amplify turbulence and ultimately confine the cosmic rays closer to the shock. The current that drives the instability is turned off either by scattering the particles driving the current or by accelerating the background plasma up to the drift speed of the escaping particles. Our simulation reveals sonic Mach numbers ranging from 5 to 300 at accretion shocks, and the Alfvénic Mach numbers would be even higher if the field were $\lesssim \!10$\,nG. An amplification factor of 3--10 can bring the upstream magnetic field to 30--1000\,nG, depending on the enrichment by early galactic outflows and subsequent turbulence.

Ongoing work on a process called the magnetic bootstrap relies on the anisotropic upstream cosmic ray flux from curved shocks to amplify the magnetic field, perhaps in a time-dependent manner \citep{2007AIPC..921...62B,2023arXiv230909116B}. In this process, the highest-energy cosmic rays stream farthest ahead of the shock and interact with waves that match their gyroradii. Rather than building up the field on smaller scales near the shock front, the highest-energy cosmic rays generate the waves needed to scatter particles of similar rigidity back toward the shock. We propose that if the high-energy end of a cosmic ray distribution is not effectively scattered by smaller-scale turbulence, an unsuppressed component will eventually dominate the far-upstream region, where its anisotropy drives instabilities to increase the scattering rate. Thus, the streaming instabilities could be most effective far ahead of the shock front, where there is an anisotropic flux of escaping particles.

The third source of magnetic field amplification is a small-scale turbulence dynamo driven by cosmic ray pressure gradients acting on density perturbations in the upstream flow. This process winds up the magnetic field on scales at and below the density perturbations \citep{2009ApJ...707.1541B,2012MNRAS.427.2308D,2014MNRAS.444..365D}\footnote{In a turbulent dynamo, the mean magnetic field does not change. Instead, magnetic energy is transferred from large to small scales.}. All that is needed are initial gas density fluctuations $\delta \rho_{\rm b}$ with an outer length scale of $\lambda_{\rho}$ and a cosmic ray pressure gradient $\nabla P_{\rm CR}$, both of which are expected upstream of any accretion shock that accelerates cosmic rays. The pressure gradient induces differential acceleration of the density fluctuations, and the resulting turbulent motions wind up and amplify preexisting magnetic fields. Magnetic energy is generated on scales $\lambda_B \leq \lambda_{\rho}$, and to effectively scatter EECRs, $\lambda_B$ must be larger than about 10\,kpc upstream of the shocks---a condition easily satisfied megaparsecs from the cluster center. We assume $\lambda_{\rho} \sim 100\text{–}500$\,kpc as a fiducial outer scale for upstream density perturbations—large enough to be set by filamentary inflows and merger driving, yet small enough to remain below the shock-curvature scale.

A simple model of this dynamo requires defining a scale height $L$ of the cosmic ray pressure\footnote{We depart slightly from the linear pressure, which is effectively $P\propto \frac{L-x}{x}$ in our notation, that is used in \citet{2012MNRAS.427.2308D} in order not to impose an artificial boundary where cosmic ray pressure is zero. With a sufficiently large scale height $L \gg \lambda_{\rho}$, this change is minor.} $P_{\rm CR} = \zeta_{\rm CR} \rho_{\rm b} u_{1}^2 e^{-x/L}$, where $u_1$ is the upstream flow speed at the shock, $x$ is the distance upstream of the shock, and $\eta$ is a positive parameter less than 1. We adopt an estimate that about 10\% of the ram pressure goes into cosmic rays, $\zeta_{\rm CR} \sim 0.1$, consistent with simulations of shock acceleration \citep{2014ApJ...783...91C}. We take the scale height to be the radius of the shock, $L \approx r_{\rm sh}$, because it is the natural scale for cosmic rays to escape from a curved shock front. The condition of a short eddy turnover time compared to the fluid crossing time of the outer scale ensures that turbulent motions can wind up the magnetic field multiple times, facilitating efficient amplification by the dynamo. This condition implies the following relation: $\lambda_{\rho} \ll \zeta_{\rm CR} \frac{\delta \rho_{\rm b}}{\rho_{\rm b}} L$, easily satisfied for accretion shocks as long as $\zeta_{\rm CR}$ is not too small. We take $\delta \rho_{\rm b} \sim \rho_{\rm b}$, supported by evidence of gas clumping in the outskirts of clusters. With the clumping factor defined as $C \equiv \frac{\langle n_{\rm e}^{2} \rangle}{\langle n_{\rm e} \rangle^2} = 1 + \frac{\sigma^2}{\langle n_{\rm e} \rangle^2}$,  \citet{2014MNRAS.437.1909M} found $C \approx 2$--3 near the virial radius.

Small-scale dynamos experience a brief exponential phase, a linear growth phase, and a saturation phase, where the magnetic energy density is similar to the kinetic energy density. The transfer rate of kinetic to magnetic energy per unit volume in the linear phase is $A_{\rm d} \rho_{\rm b} u_{1}^3 / L$, with $A_{\rm d} \approx 0.06$ \citep{2009ApJ...693.1449C,2009ApJ...707.1541B}. For an accretion shock with the outer scale $L \approx r_{\rm sh}$, the growth timescale to reach $B \sim 1\,\mu{\rm G}$ is 
\begin{align}
    \tau_{\rm mag} \sim 6 \, \rho_{\rm -29}^{-1} u_{1000}^{-3} r_{\rm Mpc} \, {\rm Gyr}, \label{eq:tau_mag}
\end{align}
using $\rho_{\rm -29} = \rho_{\rm b} / (10^{-29}\,{\rm g\,cm}^{-3})$, $u_{1000} = u_{1} / (1000\,{\rm km\,s}^{-1})$, and $r_{\rm Mpc} = r_{\rm sh}/(1\,{\rm Mpc})$. Using estimates from our simulation, Equation\,(\ref{eq:tau_mag}) yields $\tau_{\rm mag} \sim 2\,{\rm Gyr}$ for large clusters like the one in our simulated Coma analog halo. This growth rate is comparable to cluster dynamical times and satisfies $\tau_{\rm mag} \leq \tau_{\rm adv}\sim \frac{r_{\rm sh}}{u_{1}} \sim 3\,{\rm Gyr}$ for large cluster accretion shocks, ensuring that turbulence can grow before the upstream plasma is advected through the precursor. The magnetic field level upstream of the shock is limited by the advection timescale $\tau_{\rm adv}$ and by the saturation condition $e_{\rm B} \approx e_{\rm F}$. Dissipation is not relevant at the scales of turbulence that scatter cosmic rays. Following \citet{2012MNRAS.427.2308D}, the kinetic energy density is
\begin{equation}
e_{\text{F}} = \frac{1}{2} \rho_{\rm b} (\delta u)^2 \approx \frac{1}{2 \rho_{\rm b}} (\delta \rho_{\rm b})^2 \zeta_{\rm CR}^2 u_{1}^2.
\label{eq:KE-density}
\end{equation}
The saturation condition sets a limit on the strength of magnetic field
\begin{equation}
  B \leq 2 \pi^{1/2} \frac{ (\delta \rho_{\rm b})}{\rho_{\rm b}^{1/2}} \zeta_{\rm CR} u_{1} \approx 0.1 \, \zeta_{\rm 0.1} \rho_{-29}^{1/2} \,u_{1000} \, \mu{\rm G},
  \label{eq:B-dynamo}
\end{equation}
using $\delta \rho_{\rm b} \sim \rho_{\rm b}$. We expect the gas density to be roughly $2 \times 10^{-29}\,{\rm g\,cm}^{-3}$, and the clumping factor may be $C\gtrsim 3$, implying $\delta \rho_{\rm b} > 2 \rho_{\rm b}$. Favorable, but still plausible, values for $C$, $\rho_{\rm b}$, $\zeta_{\rm CR}$, and $u_{1}$ can get the magnetic field close to the target value of 1\,$\mu$G.

As noted by \citet{2009ApJ...707.1541B}, the turbulent magnetic field generated by the small-scale dynamo suppresses streaming instabilities, which require a smooth background field in the linear regime. Therefore, the streaming instability is unlikely to be important at the same time and place as the small-scale dynamo. However, it is more likely to act both in the early stages of shock formation and far upstream, where the escaping cosmic rays maintain an anisotropic flux in a less turbulent background. During these early stages of magnetic amplification, the first generation of escaping cosmic rays develops the pressure gradient necessary for the dynamo to operate. In addition, the streaming instability generates density fluctuations on the scale resonant with the highest-energy particles, augmenting preexisting inhomogeneities of infalling matter. These combined fluctuations may enhance the clumping factor upstream compared to measurements downstream on the outskirts of clusters. In doing so, the streaming instability provides the necessary conditions for the dynamo to further amplify magnetic fields.

Cosmological MHD simulations show that large-scale compression and shear alone can amplify magnetic fields in cluster outskirts and filaments to $B\sim10\!-\!100\,{\rm nG}$, even without plasma instabilities or small-scale dynamos \citep{2019A&A...627A...5V}. We argue that the combination of galactic outflows, the current-driven streaming instability, and the pressure-driven small-scale dynamo can then sequentially step up the field to microgauss levels. The conditions for magnetic field amplification vary in both space and time, and the relative importance of the streaming instability and the kinetic dynamo will likewise vary.

\subsection{Factors determining the maximum cosmic ray energy}
\label{subsec:maximum-energy}

The observed cosmic ray spectrum exhibits such a steep decline above 50\,EeV that the question of maximum energy, $E_{\rm max}$, really has two parts. First, can accretion shocks routinely accelerate particles up to around 50\,EeV? Second, under exceptionally favorable conditions---such as in particularly massive clusters or with unusually strong shocks---can these shocks push particles to extreme energies near 200\,EeV?

We characterize the limiting factors for $E_{\rm max}$ with six key timescales: the growth timescale of magnetic turbulence ($\tau_{\rm mag}$), the advection timescale ($\tau_{\rm adv}$), the particle acceleration timescale ($\tau_{\rm acc}$), the diffusion timescale from the shock ($\tau_{\rm diff}$), the energy-loss timescale ($\tau_{\rm loss}$), and the shock lifetime ($t_{\rm sh}$).

The growth timescale of magnetic turbulence ($\tau_{\rm mag}$) depends on the specific mechanism for the growth of magnetic fields. For our model to be valid, $\tau_{\rm mag} \leq \tau_{\rm adv} \sim \frac{r_{\rm sh}}{u_{1}} \sim (2$--$3)\,{\rm Gyr}$. In Section\,\ref{subsec:magnetic-fields} we assumed that 6\% of the upstream flow kinetic energy is continuously transferred to magnetic energy via a turbulent dynamo, and that rate is sufficient for fast shocks---with $\tau_{\rm mag}/\tau_{\rm adv} \propto u_{1}^{-2}$---to satisfy this criterion. Alternatively, one can say the magnetic field is limited by the advection time: $B^2 \leq 8\pi A_{\rm d} \rho_{\rm b} u^2 $, or $B \lesssim 400\,\rho_{-29} u_{1000}^2\,{\rm nG}$. This $B \propto u_{1}^2$ dependence is important for the acceleration timescale.

According to DSA theory, the acceleration timescale $\tau_{\rm acc }$ for a nonrelativistic shock is
\begin{align}
   \tau_{\rm acc} = \frac{p}{\frac{\mathrm{d} p}{\mathrm{d}t}} = \frac{3}{u_1 - u_2} \left( \frac{D_1}{u_1} + \frac{D_2}{u_2} \right),
\end{align}
where $u_1$ and $u_2$ denote the upstream and downstream flow speeds relative to the shock, respectively, and $D_1$ and $D_2$ are the corresponding diffusion coefficients \citep{1983RPPh...46..973D}. For all accretion shocks with $M \geq 5$, as we have selected, the standard compression ratio is $r \equiv u_1/u_2 = 4$.
Assuming that the downstream compression of the magnetic field reduces the diffusion coefficient by the same factor $r$,  we have $D_1 \approx r D_2$ and obtain a common form of the acceleration timescale \citep{2013APh....43...56B,2014ApJ...794...47C}:
\begin{align}
    \tau_{\rm acc } = \frac{8 D_1}{u_1^2}.
    \label{eq:tau_accel-1}
\end{align}
A more accurate prescription of particle diffusion would allow the diffusion coefficient to vary with rigidity and distance from the shock, expressed as $D(R,r)$. In that case, high-rigidity particles may spend far more time upstream than downstream, potentially modifying Equation\,(\ref{eq:tau_accel-1}). However, for simplicity, we proceed with this standard treatment.

Hybrid simulations (kinetic ions and fluid electrons) \citep{2014ApJ...794...47C} suggest that cosmic rays undergo Bohm diffusion, where the diffusion coefficient at rigidity $R$ is $D_B(R) = \frac{r_{\rm g} c}{3}$, with $r_{\rm g} = \frac{p}{qB}$ as the gyroradius in the amplified magnetic field. In their simulations, \citet{2014ApJ...794...47C} found that the ratio $\kappa \equiv D(R) / D_B(R) \approx 1.2$ for shocks with Mach number $M = 60$, suggesting Bohm-like diffusion is a reasonable approximation. We therefore adopt $\kappa = 1$ for the strongest shocks, which accelerate the highest-rigidity particles.

Assuming Bohm-like diffusion, the acceleration timescale is
\begin{eqnarray}
    \tau_{\rm acc} \approx 850  \, \frac{R_{18} }{Z\, B_{\mu{\rm G}} \, u_{1000}^{2} } \, {\rm Myr},
    \label{eq:tau_accel-numeric}
\end{eqnarray}
where $R_{18} = R / (10^{18}\,{\rm V})$ and $B_{\mu{\rm G}} = B/(1\,\mu{\rm G})$. As noted earlier, if the magnetic field strength scales as $B \propto u_{1}^2$, limited by the upstream advection time, then the acceleration timescale follows $\tau_{\rm acc } \propto u_1^{-4}$. This strong dependence on flow speed\footnote{Here, we assume the accretion shock is stationary so that the upstream reference frame coincides with the cluster rest frame. During the shock's evolution, it must have experienced outward shock speeds as high as $v_{\rm sh} = u_2 = u_1 /4$ as the co-moving size of the cluster grew. In that case, infalling gas with velocity $u'_1$ in the cluster frame has a speed relative to the shock of $u_1 = u'_1 + v_{\rm sh} = (4/3) u'_1$.} implies that only regions with the fastest accretion shocks can accelerate the highest-rigidity cosmic rays.

For particles to reach the highest rigidities, the acceleration timescale must be shorter than three competing timescales: the shock lifetime, which limits the duration over which acceleration can occur; the loss timescale, which accounts for energy losses through interactions with the background radiation; and the escape timescale, which represents the time required for particles to diffuse away from the shock region. 

Regarding the shock lifetime, our simulation shows strong shocks existing at $z=1$ and increasingly weaker shocks further back in time. Assuming that strong shocks have been in continuous existence since $z=1$ yields an available shock lifetime of $t_{\rm sh} \sim 8$\,Gyr. Assuming that the accelerating to the highest rigidities can begin only after sufficient time for the magnetic field to amplify, one condition on the maximum energy is that $\tau_{\rm acc} \leq t_{\rm sh} - \tau_{\rm mag} \sim (5$--$6)\,{\rm Gyr}$. 

During acceleration, the energy-loss limit is determined by the balance between the acceleration timescale $\tau_{\rm acc}$ and the loss timescale $\tau_{\rm loss}$ due to interactions with the cosmic microwave background (CMB) and the diffuse extragalactic background light (EBL). These losses include Bethe-Heitler pair production, pion photoproduction, and photodisintegration for heavy nuclei. The suppression of UHECR flux due to these effects is commonly referred to as the Greisen–Zatsepin–Kuzmin (GZK) effect \citep{1966PhRvL..16..748G,1966JETPL...4...78Z}.

\begin{figure}[htb!]
    \centering
    \includegraphics[width=\columnwidth]{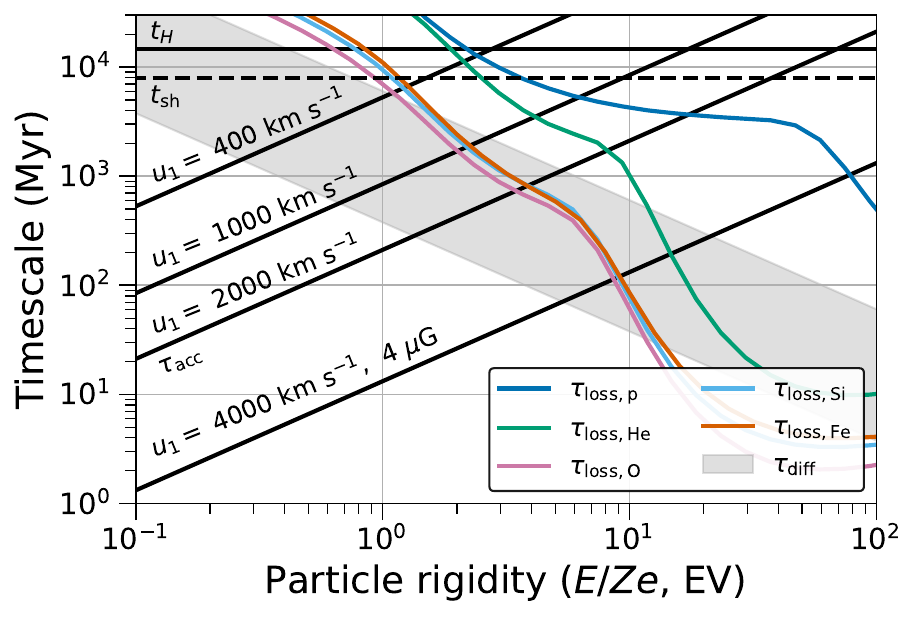}
    \caption{Loss timescales (colored curves) as a function of particle rigidity for different species: protons, helium, oxygen, silicon, iron. Adiabatic (Hubble) expansion losses are labeled with $t_H$. The acceleration timescales as a function of rigidity following Equation\,(\ref{eq:tau_accel-numeric}) are plotted as black solid lines with different values of $u_{1}$ (400, 1000, 2000, and 4000\,km\,s$^{-1}$). The upstream amplified magnetic field at the length scales resonant with the cosmic rays is assumed to be $B_{\rm tot} = 1\,\mu{\rm G}$, except for one line labeled with 4\,$\mu$G. The shaded band indicates the diffusion timescale, $\tau_{\rm diff}$, for $L_{\rm esc}=0.5$–2\,Mpc assuming uniform magnetic field. The dashed line is the shock lifetime, $t_{\rm sh}$. The crossing points of $\tau_{\rm acc}$ with $\tau_{\rm loss}$ marks the maximum rigidity allowed for that species with the specified shock parameters.}
    \label{fig:loss_accel_v_rig}
\end{figure}

We use the loss lengths $\chi_{{\rm loss}} \equiv c \,\tau_{\rm loss} \equiv -cE / (dE/dt)$ as presented by \citet{2006JCAP...09..005A}. Instead of treating every species of nuclei separately, we use protons, helium, oxygen, silicon, and iron as proxies for cosmic rays of similar atomic mass. When plotted as a function of rigidity in Figure~\ref{fig:loss_accel_v_rig}, $\tau_{{\rm loss}}(R)$ for oxygen, silicon, and iron are very similar to each other, suggesting this crude categorization is adequate for our purposes. The adiabatic expansion of the Universe provides a loss timescale equal to the Hubble time (horizontal solid line in the figure). For reference, $t_{\rm sh}$ is plotted as the elapsed cosmic time since $z= 1$ (dash-dotted line), calculated using the online calculator from \citet{2006PASP..118.1711W} and the same parameters as in Section\,\ref{sec:sims}. 

Figure\,\ref{fig:loss_accel_v_rig} shows the crossing points of the loss timescale curves and the acceleration timescale curves for different values of $u_1$, with $B_{\rm tot} = 1\,\mu\mathrm{G}$. These crossing points determine the rigidity cutoff achievable by particles in shocks with varying speeds when energy losses provide the most stringent constraint. We plot the acceleration timescales for three representative values of $u_1$: a relatively low speed typical of filament shocks ($u_1 = 400\,\mathrm{km\,s}^{-1}$), a characteristic cluster flow speed ($u_1 = 1000\,\mathrm{km\,s}^{-1}$), and a high speed expected outside the largest clusters ($u_1 = 2000\,\mathrm{km\,s}^{-1}$). For more typical cluster shocks, the rigidity cutoff, $R_{\rm cut}$, is approximately 2\,EV for heavy nuclei, which corresponds to $50\,\mathrm{EeV}$ for iron. Protons, however, have a higher rigidity cutoff but are unlikely to exceed $7\,\mathrm{EeV}$, except under very favorable shock conditions.

In most scenarios, particle escape limits the maximum rigidity more than energy losses. In accretion shocks, cosmic rays can escape either by advection downstream or by diffusion upstream, and the relative importance of the two processes depends on rigidity. At low rigidities, advection downstream dominates. Higher rigidity particles can diffuse farther from the shock front where the scattering is weaker, and the diffusive escape upstream ultimately limits the maximum rigidity. A full treatment would require modeling the escape timescale $\tau_{\rm esc}^{-1} = \tau_{\rm adv}^{-1} + \tau_{\rm diff}^{-1}$ with a diffusion coefficient that varies with both rigidity and distance from the shock. For simplicity, we assume uniform Bohm diffusion, giving a characteristic diffusion timescale $\tau_{\rm diff} \simeq L_{\rm esc}^2/6D$, where $L_{\rm esc}$ is the characteristic upstream escape length, typically comparable to the shock radius. Figure\,\ref{fig:loss_accel_v_rig} shows $\tau_{\rm diff} \approx 1.5\, L_{\rm Mpc}^2 B_{\mu{\rm G}} R_{\rm EV}^{-1}\,{\rm Gyr}$ as a shaded band for $L_{\rm Mpc} = L_{\rm esc}/(1\,{\rm Mpc})$ ranging from 0.5 to 2. Setting $\tau_{\rm acc} = \tau_{\rm diff}$ yields a rigidity cutoff:
\begin{align}
    R_{\rm cut}^{\rm(diff)} \;&\simeq\; \frac{3}{\sqrt{48}}\,\frac{u_1}{c}\,B\,L_{\rm esc} \;\simeq 1 \, B_{\mu{\rm G}} \, u_{1000} \, L_{\rm Mpc} \,{\rm EV}.
    \label{eq:rigidity_cutoff}
\end{align}
This scaling is only a factor of about 3 lower than the standard Hillas criterion based on geometric confinement \citep{1984ARA&A..22..425H}. For particles that reach Earth from distant clusters, the diffusive escape timescale must be shorter than the energy-loss timescale ($\tau_{\rm diff} < \tau_{\rm loss}$); otherwise they would lose energy through pair production or photodisintegration before escaping. For massive clusters, this diffusive escape limit yields $R_{\rm cut}^{\rm(diff)} \approx 1–2\,{\rm EV}$ for all species. Again, this limits iron nuclei to $\lesssim 50$\,EeV for typical conditions. For comparison, accretion shocks around thin filaments are constrained by this condition, giving a maximum rigidity below $\sim$1\,EV.

The rigidity cutoff for each species depends on whether the diffusion or energy-loss timescale is shorter. For $\tau_{\rm diff}$ within the shaded band in Figure\,\ref{fig:loss_accel_v_rig}, $R_{\rm cut}^{({\rm diff})}$ for all species ranges from about 700\,PV to 3\,EV assuming fiducial values of $u_{1000} = 1$ and $B_{\mu{\rm G}} = 1$. By contrast, the energy-loss cutoff $R_{\rm cut}^{({\rm loss})}$ for protons is about 6\,EV (and slightly less for helium), indicating that light nuclei are diffusion-limited. For all other nuclei, $R_{\rm cut}^{({\rm loss})} \approx 2\,{\rm EV}$, so energy losses and escape are comparable.

This diffusion-limited escape not only sets the maximum rigidity but also determines the shape of the high-energy cutoff. The familiar exponential cutoff arises when escape is energy-independent (e.g., finite age or advection downstream). By contrast, for diffusion-limited escape the escape time decreases with rigidity, $\tau_{\rm diff}\propto R^{-1}$. Inserting this escape timescale into the standard DSA solution yields a super-exponential cutoff, $N(R)\propto R^{-q}\exp[-(R/R_{\rm max})^{2}]$. \citet{2024ApJ...977L..18C} have argued that to explain the small dispersion of $X_{\rm max}$ at a given energy, the UHECR spectrum needs either an unusually hard spectral index or a cutoff sharper than exponential. Cluster accretion shocks can be consistent with the UHECR spectrum with a normal spectral index $s\sim 2.0$--$2.2$ and a super-exponential cutoff.

To explore the upper limits of this model to explain EECRs, we include a maximally optimistic acceleration timescale curve for $u_1 = 4000\,\mathrm{km\,s}^{-1}$ and $B_{\rm tot} = 4\,\mu\mathrm{G}$. Such extreme conditions may occur only rarely but provide a useful benchmark for the theoretical limits of shock acceleration in these environments. Under these parameters, the rigidity cutoff is about $5\!-\!20\,\mathrm{EV}$ for protons and $5\!-\!9\,\mathrm{EV}$ for all other nuclei, depending on the escape length.
The Auger data through December 2020 has 2635 events above 32\,EeV but only 35 events ($\sim1$\%) above 100\,EeV \citep{2022ApJ...935..170A}. We contend that events such as the Amaterasu event---with an energy of $\mathord{\sim}240\,{\rm EeV}$ \citep{2023Sci...382..903T}---can be explained by iron nuclei accelerated to $R\sim 9\,{\rm EV}$. This extreme rigidity is possible either from an exceptionally favorable region of the shock front or from the super-exponential tail of a shock with $R_{\rm cut} \sim 5\,{\rm EV}$. Alternatively, such EECRs consist of even heavier nuclei (e.g., lead with $Z = 82$), which are present in the Galactic CR spectrum and could be reaccelerated at extragalactic shocks. However, their low abundances make this possibility highly speculative.

After escaping the shock, cosmic rays must traverse intergalactic space where the GZK effect reduces their energy, while diffusion through extragalactic magnetic fields affects the arrival directions and delays their propagation. A general statement of the often-quoted GZK limit is $t_{\rm prop} \lesssim \tau_{\rm loss}$, where $t_{\rm prop}$ is the propagation time for an individual cosmic ray. The stochastic nature of propagation and energy losses means that some particles will propagate longer than the energy-loss timescale for a given rigidity at the cost of exponentially reduced flux and diminishing mass number, as heavy nuclei lose nucleons through the giant dipole resonance and quasi-deuteron processes. EECRs observed at Earth most likely originated within $\sim$40\,Mpc to avoid severe losses, whereas lower-energy UHECRs have a much larger cosmic ray horizon, which depends on both energy losses and diffusion in intergalactic magnetic fields.

 We can summarize all of the timescale criteria in this subsection as follows:
\begin{subequations}
\begin{align}
    \tau_{\rm mag} &\leq \tau_{\rm adv} \\
    \tau_{\rm esc}^{-1} &= \tau_{\rm adv}^{-1} + \tau_{\rm diff}^{-1} \\
    \tau_{\rm acc} &\leq \min( \tau_{\rm loss}, \tau_{\rm esc}, t_{\rm sh} - \tau_{\rm mag})
\end{align}
\end{subequations}
Each of these relations depends on rigidity and particle species. The entire sequence of magnetic field amplification, particle acceleration, and propagation must fit within a Hubble time, $t_H$. The maximum energy of UHECRs depends on the interplay of these timescales. 

\section{A hierarchical model of shock acceleration}
\label{sec:hierarchcal-model}

The new contribution we present to the framework of accretion-shock acceleration of cosmic rays is the multiscale hierarchy of shocks. We view the relatively smooth cosmic ray spectrum as an indication that the same acceleration process---DSA---operates at all scales. The observed smooth transition between Galactic and extragalactic cosmic rays is remarkable given the necessity of multiple sources, each contributing spectral breaks due to rigidity limits, propagation effects, or source-class transitions. This smoothness suggests that cosmic rays escaping from galaxies continue their acceleration in larger extragalactic shocks. Thus, understanding the source of the highest-energy cosmic rays requires understanding how galactic\footnote{Here, ``galactic'' refers to cosmic rays from any galaxy, not just Galactic cosmic rays from our own Galaxy.} cosmic rays escape their home galaxies, propagate to the next level of shocks, and accelerate to higher energy.

Typical star-forming galaxies accelerate cosmic rays at least above 1\,PV, and the higher-energy cosmic rays escape in less than 10\,Myr. As detailed in our previous work (\citealt{2023ApJ...953...49M}), galaxies such as our own exhibit modest galactic winds that can reach over 800\,km\,s$^{-1}$ at a radius of about 200\,kpc, where the outflow forms a shock against the pressure of the intergalactic medium. The escaping galactic cosmic rays inevitably reach the shocks surrounding their home galaxies and are reaccelerated to $10$--$40\,{\rm PV}$. The acceleration process is the same as that at a supernova remnant shock, but the geometry is inverted. For particles to return to their home galaxy, they must escape upstream against the galactic wind. The upstream flux would appear at most as an enhancement spanning less than a decade in energy. For this reason, the returning flux from our Galactic wind termination shock is unlikely to account for a large fraction of the observed spectrum at Earth, but the downstream flux that escapes our Galaxy continues to propagate outward until it reaches the next shock in the hierarchy. Starburst galaxies with stronger galactic winds have stronger shocks that could accelerate cosmic rays to even higher energies, but they are unable to explain EECRs.

Since most galaxies emitting cosmic rays are contained within the large-scale filamentary structure, escaping cosmic rays above 1\,PV need only travel a few megaparsecs to reach an accretion shock, likely a filament shock, which is the most abundant type of accretion shock by surface area. While diffusing in magnetic turbulence below 10\,nG at resonant scales, these cosmic rays can reach these shocks within a few billion years. Filament shocks, in turn, can accelerate petavolt cosmic rays up to 100--1000\,PV over similar timescales. These exavolt cosmic rays then escape those shocks and may eventually reach cluster accretion shocks, which can continue the acceleration to the highest rigidity, about 2--9\,EV, which corresponds to 52--230\,EeV for iron nuclei, depending on the properties of the cluster. Alternatively, cluster accretion shocks can accelerate particles directly from the lower-energy population of cosmic rays escaping galaxies nearby and within the cluster.

One major consequence of the accretion shock model is that if our Galaxy resides within a filamentary or sheet-like shock structure, the observed cosmic ray spectrum will exhibit
a transition from a soft to a hard spectral index as the flux changes from predominantly downstream emission from local filament shocks to upstream emission from more distant cluster shocks \citep{2014PhDT.......656S}. The local filamentary structure in which the Local Group is contained acts as a reservoir for the downstream cosmic rays accelerated by the filament shocks. These cosmic rays escape into the surrounding voids in a rigidity-dependent way, which steepens the spectral index below the ankle by $\delta_{\rm leak} \simeq 0.3\!-\!0.6$. In addition, large accretion shocks exhibit spatial and temporal variations in $u_1$ and $B$, producing a spread in rigidity cutoffs that likely follow a rigidity dependence $R_{\rm cut}^{-\delta_{\rm source}}$. If $\delta_{\rm source} \simeq 0.5\!-\!0.8$, then the combined softening of these two effects yields a spectral index $s_{\rm obs} \approx s + \delta_{\rm leak} + \delta_{\rm source} \approx 3.3$, consistent with the observed spectrum below the ankle.

The hardening at the ankle marks the transition to UHECRs that have escaped from strong cluster shocks, which do not exhibit the spectral softening seen in the downstream flux of the filaments. Although cluster shocks have a range of rigidity cutoffs, the observed flux above the ankle is dominated by the most massive and luminous clusters, whose cutoffs approach the limits imposed by energy losses. Above the ankle, the observed spectral index is closer to the source value of $s \simeq 2.2$. This natural division between the softer flux from local filaments and the harder flux from clusters yields two distinct components of extragalactic cosmic rays.

We expect this spectral transition to occur around the maximum rigidity for filament shocks. From the timescales shown in Figure~\ref{fig:loss_accel_v_rig}, protons and helium reach a maximum rigidity of about 1–3\,EV at filament shocks, depending on the shock conditions. Below the spectral transition---around 5\,EeV, where Auger observes the ‘ankle’---we propose that the spectrum is dominated by protons and helium accelerated and advected downstream from filamentary shocks surrounding the Local Group or Local Sheet. Above the spectral transition, the hard cluster-shock spectrum with a super-exponential cutoff explains the rapid increase in average mass due to the rigidity break around 2–4\,EV for medium and heavy nuclei.


These features are consistent with the best-fit models of the Pierre Auger Collaboration \citep{2023JCAP...05..024A}, which include a low-energy component to explain the mixed composition below the ankle, sometimes interpreted as a Galactic contribution with an unusually high rigidity cutoff. In our framework, this component arises naturally from medium-mass nuclei accelerated in filament shocks, eliminating the need for a separate Galactic population with a high rigidity cutoff.  

The presence of both filamentary and cluster shocks in the hierarchical model not only explains the observed spectral transition but also increases the total power available for cosmic ray production. Filament shocks, with their extensive surface area, contribute between 10\% and 50\% of the luminosity of all of the shocks in the universe with $M > 5$ and $\delta < 1000$. In Section\,\ref{sec:sim_results}, we reported this average total shock power density of $\mathcal{P}_{\rm LSSS} \approx 3.4 \times 10^{-35}$~W~m$^{-3}$, compared to the oft-quoted value for the required luminosity density of the sources of UHECRs above 1\,EeV, $\mathcal{L}_{\rm CR}(>1\,{\rm EeV}) \sim 10^{-38}\,{\rm W\,m}^{-3}$. The luminosity density requirements of the two scenarios of extragalactic cosmic rays are each $\sim\! 10^{-37}\,{\rm W\,m}^{-3}$ \citep{2023JCAP...05..024A}, which is about $3\times 10^{-3} \,\mathcal{P}_{\rm LSSS}$. Other estimates for the required UHECR luminosity density range between $10^{-37}\,{\rm W\,m}^{-3}$ and $5\times 10^{-35}\,{\rm W\,m}^{-3}$, depending on the source spectral index and the energy range \citep{2003ehcr.conf...41S}.

The hierarchical model of accretion shocks also predicts a non-uniform spatial distribution of cosmic ray energy density. Cosmic rays below about 1\,EV are predominantly confined within the filamentary structure of the large-scale universe, as these particles lack the rigidity to escape upstream of the filament shocks. This feature further lowers the power requirements of the shocks because they do not have to supply the voids with the elevated cosmic ray energy density we observe within the structure. Conversely, cosmic rays above about 1\,EV may fill intergalactic space but with a very hard spectrum that does not extend down to $10^{15}$\,eV or $10^{17}$\,eV as some models propose. Again, this feature lowers the power requirements of cluster shocks.

\begin{table*}[htbp!]
\centering
\caption{Nearby Large Clusters in the Local Universe. The masses listed are $M_{200}$, except for Norma, for which we list the dynamical mass within 2\,Mpc of the center of the cluster. Mass estimates for clusters within the Pavo--Indus Supercluster are currently unavailable. $E_{\rm cut,Fe}^{\rm(diff)}$ is the estimated cutoff energy for iron from each cluster using Eq.~\ref{eq:rigidity_cutoff}.}
\begin{tabular}{lcccc}
\hline
\textbf{Cluster Name} & \textbf{Distance (Mpc)} & \textbf{Mass ($10^{14} \, M_\odot$)} & \textbf{Supercluster Association} & $E_{\rm cut, Fe}^{\rm(diff)}$ (EeV) \\
\hline
Virgo Cluster & 16 & \(6 \) & Virgo & 10 \\
Fornax Cluster & 20 & \( 1  \) & Virgo & 2 \\
Antlia Cluster & 40 & \( 3 \) & Hydra–Centaurus & 5 \\
Centaurus (Abell 3526) & 47 & \( 2 \) & Hydra–Centaurus & 3 \\
Hydra I (Abell 1060) & 50 & \( 2 \) & Hydra–Centaurus & 3 \\
Norma (Abell 3627) & 70 & \( 10 \) & Hydra–Centaurus & 20 \\
Pavo I (Abell S0805) & 70 & \( - \) & Pavo--Indus & \( - \) \\
Pavo II (Abell S0806) & 70 & \( - \) & Pavo--Indus & \( - \) \\
Perseus (Abell 426) & 80 & \( 7 \) & Perseus–Pisces & 10 \\
Leo (Abell 1367) & 90 & \( 7 \) & Coma & 10 \\
Coma (Abell 1656) & 100 & \( 10 \) & Coma & 20 \\
\hline
\end{tabular}
\label{tab:clusters}
\end{table*}

We have focused on galaxy clusters and the filaments between them within the cosmic ray horizon that may accelerate UHECRs up to 200\,EeV, but these clusters themselves often form large-scale superclusters in various configurations of walls, filaments, or nodes. Such superclusters are not gravitationally bound, but the combined effect of denser and more focused flows of matter onto clusters could result in faster and stronger shocks around the individual clusters. Moreover, the proximity of multiple accretion shocks around nearby clusters increases the UHECR luminosity from these superclusters. Table\,\ref{tab:clusters} lists some of the largest clusters within about 100\,Mpc. Cosmic rays from the edge of this cosmic ray horizon may not reach our Galaxy at the highest energies. References for Table\,\ref{tab:clusters}: Virgo \citep{2020A&A...635A.135K}, Fornax \citep{2001ApJ...548L.139D}, Antlia \citep{2015MNRAS.452.1617H}, Centaurus \citep{2013MNRAS.432..554W}, Hydra I \citep{2022A&A...659A..92L}, Norma \citep{2008MNRAS.383..445W}, Perseus \citep{2011Sci...331.1576S}, Leo \citep{1998ApJ...505...74G}, Coma \citep{2007ApJ...671.1466K}.

We estimate the cutoff energy for each cluster using the maximum confinement rigidity given in Equation\,(\ref{eq:rigidity_cutoff}) because diffusive escape is typically more restrictive than energy losses for the cluster parameters considered here. We take the escape length to be the virial radius and use the saturation level of the magnetic field from the small-scale dynamo given in Equation\,(\ref{eq:B-dynamo}). The scalings for the virial radius and flow speed are from Section\,\ref{sec:sim_results}. With $\zeta_{\rm CR} = 0.1$ and $\rho_b = 2 \times 10^{-29}\,{\rm g\,cm}^{-3}$, we find $R_{\rm cut} \sim 0.06 \,M_{14}\,{\rm EV}$. Table\,\ref{tab:clusters} lists the energy cutoff using iron, $E_{\rm cut,Fe} \approx 1.6 \,M_{14}\,{\rm EeV}$. Many approximations go into this estimate, and, as mentioned in Section\,\ref{subsec:maximum-energy}, exceptional regions of the shock front, the super-exponential tail of the distribution, and trans-iron elements could yield energies above this energy cutoff. These values should therefore be regarded as estimates rather than strict upper limits, especially if our adopted scaling for $u_1$ underestimates the true flow speed.

\section{Observational signatures}
\label{sec:observations}

\subsection{Cosmic ray observables}

The three reconstructed elements of indirect-detection experiments are the energy spectrum, the composition, and the arrival direction. We argued in Section~\ref{sec:hierarchcal-model} that the hierarchical model of shocks naturally explains the general features of the spectrum and composition. While the detection of cosmic rays with energy $>$300\,EeV (or protons with energy $>$80\,EeV) could rule out iron accelerated at accretion shocks, further refinement of the spectral shape and composition are unlikely to distinguish models without a better understanding of the sources and propagation.

The remaining cosmic ray observable, the arrival direction, favors sources distributed with the large-scale structure of the universe. No experiment has found a statistically significant source, confounded by the deflection of the intervening magnetic fields. After accounting for the deflection of the Galactic magnetic field, the Amaterasu particle's \citep{2023Sci...382..903T} arrival direction points to a region of the local universe without an obvious source \citep{2024ApJ...962L...5U}. However, a strong accretion shock separates the Local Group from the Local Void, and we have no model for this extragalactic deflection. The average propagation distance for a 244\,EeV particle is $\sim$30\,Mpc for both protons and iron \citep{2023Sci...382..903T}, long enough for a cosmic ray from the Virgo Cluster to scatter off a nearby accretion shock, changing its arrival direction.

The only statistically significant measurement of dipole anisotropy for UHECRs is 7\% above 8\,EeV \citep{2024icrc.confE.521T}. Comparisons with the anisotropy of gamma-ray sources indicate that the sources of UHECRs must be relatively dim in gamma rays in order to keep the anisotropy low \citep{2024ApJ...967L..15P}. We suggest that the low anisotropy of UHECRs is due to the diffusion of cosmic rays from a few extended sources in the local universe. The large-scale structure---influenced especially by the Virgo Cluster, the closest large cluster---naturally produces a cosmic ray dipole of the right amplitude \citep{2017ApJ...839L..22G}. While many types of sources generally trace the large-scale structure, accretion shocks around the cosmic web follow it more directly than other sources that can be influenced by feedback and environmental factors.

\subsection{Constraints on accretion shock suitability for accelerating UHECRs}
\label{subsec:obs_constraints}

Given the limitations of cosmic ray observables, secondary signatures are invaluable for testing models of UHECR origin. Accretion shocks exhibit multimessenger and multiwavelength secondary processes whose extended and steady nature contrasts with compact sources such as AGNs, gamma-ray bursts, or neutron stars. Inelastic collisions of cosmic rays with the intra-cluster medium (ICM) generate neutral and charged pions. The neutral pions decay into gamma rays, while the charged pions decay into neutrinos and secondary electrons. These secondary electrons generate radio synchrotron radiation and gamma rays through inverse Compton scattering off the CMB. In addition, primary electrons directly accelerated at accretion shocks also produce synchrotron and inverse Compton emission. The primary electrons are important for radio signatures, and pion decay is the most important process for gamma-ray observables. See \citet{2010MNRAS.409..449P} for an overview of these processes and the gamma-ray observables of galaxy clusters. Future observations or upper limits of neutrinos in the exavolt range would directly probe the UHECRs themselves. Without those experiments, the lower-energy secondaries probe the environments around accretion shocks and their suitability for accelerating 100\,EeV cosmic rays.

Energy losses clearly limit the maximum energy of electrons rather than acceleration time or confinement. The ratio of synchrotron losses to inverse Compton losses is $\frac{P_{\rm synch}}{P_{\rm IC}} = \frac{U_B}{U_{\rm ph}} \approx 0.1 B_{\mu{\rm G}}^2$ for the CMB background. The shock front compresses the downstream magnetic field by a factor\footnote{Only the perpendicular component of the magnetic field is compressed, so the total downstream magnetic field is $B_{2} = B_{1} \sqrt{\frac{1}{3} + r^2 \frac{2}{3}}$.} of $\sqrt{11} \,B_{1}$ for an upstream magnetic field of 1\,$\mu$G, assuming a standard compression ratio of $r=4$. Immediately downstream of the shock, the synchrotron and inverse Compton losses are comparable, and the loss timescale is at most the inverse Compton timescale of $\tau_{\rm IC} \approx 1\,E_{\rm GeV}^{-1}\,{\rm Gyr}$. Setting this loss timescale equal to the acceleration timescale yields a maximum electron energy of $E_{\rm e,max} \approx 40\, u_{1000}\,{\rm TeV}$, and the loss time at $E_{\rm e,max}$ is below 100\,kyr. The Klein--Nishina effect becomes important around this energy, but it does not affect these numbers much \citep{2009A&A...497...17V}. The turbulent decay of the magnetic field at the sub-parsec scale, relevant for these electrons, will be much faster---$\tau_{\rm decay} \sim \frac{L}{v_{\rm turb}} \sim \frac{0.1\,{\rm pc}}{10\,{\rm km\,s}^{-1}} \sim 10\,{\rm kyr}$---than the synchrotron-loss timescale. The magnetic turbulence on these small scales will decay down to $\sim$1\,$\mu$G, which we assume is supported both upstream and downstream by processes mentioned in Section\,\ref{subsec:magnetic-fields}. Therefore, the synchrotron emission will appear as a bright, thin shell immediately behind the shock, where the magnetic field is the highest. Faint, diffuse synchrotron and inverse Compton emission extend farther downstream and may have a softer spectrum from radiative cooling.

To date, observational evidence of accretion shocks is sparse but growing. \citet{2022ApJ...933..218B} detected what they called an accretion relic at about 3.5\,Mpc from the center of the Coma Cluster, but the observation lacks a spectral index to confirm it is from a hard electron spectrum. \citet{2023ApJ...943..119H} modeled the synchrotron emission from shock-accelerated electrons and found it consistent with the radio observations. \citet{2018ApJ...869...53K} found coincident, elongated rings around the Coma Cluster in $\sim$0.1\,keV X-rays, $>$GeV gamma rays, and $\sim$220\,GeV gamma rays from \textit{ROSAT}, \textit{Fermi}-LAT, and VERITAS, respectively. They find the signal consistent with a virial shock depositing $\sim$0.3\% of its energy over a Hubble time and an electron spectral index of $p\simeq 2.0$–$2.2$. \citet{2019A&A...622A.136H} claim the first detection of a virial shock with \textit{Planck} thermal Sunyaev-Zel'dovich data.

Stacked analyses of many clusters across radio, X-ray, and gamma-ray bands detect similar excess signals, all consistent with a flat electron spectral index, a cosmic ray electron (CRe) injection rate of $\sim$0.6\%, and a virial shock radius of $\sim\!2.5 R_{500} \approx 1.6 R_{200}$ \citep{2018JCAP...10..010R,2023MNRAS.521.5786H,2024A&A...686L..16I,2024JCAP...10..008I}. This radius is consistent with the accretion shock radius of our simulation. The CRe injection rate, $\dot{m} \xi_{\rm e}$, is consistent with the product of the standard cosmic ray energy efficiency, $\eta_{\rm CR} \approx 0.1$, and an electron-to-proton energy ratio of $\kappa_{\rm ep} \approx 0.05$, for $q_{\rm e} = q_{\rm p} = 2.2$ from \citet{2014A&A...567A.101P}. \citet{2023MNRAS.521.5786H} estimate the post-shock magnetic field from the stacked radio analysis with two approaches to break the degeneracy of the accretion rate, electron energy, and magnetic energy: (1) adopt $\dot{m} \xi_{\rm e} \simeq 0.006$ from gamma-ray observations and (2) assume equipartition between the electron and magnetic energy densities. Depending on the assumptions, the estimates of the fraction of shock energy converted to magnetic energy ranges from 1\% to 8\%, implying a magnetic field of 0.1–0.6\,$\mu$G. This range is slightly lower than our fiducial value of 1\,$\mu$G, but this value represents an average over the whole shock for all 44 clusters in the stacked sample. The region $\sim$10\,kpc downstream of the shock could have higher fields on smaller scales. As mentioned earlier, accretion shocks undoubtedly have some regions that have higher-than-average flow speeds and magnetic fields for the production of the highest-rigidity cosmic rays. 

As for filaments, \citet{2023SciA....9E7233V} stacked pairs of physically nearby clusters to look for residual emission from filaments. They found highly polarized radio emission, $\geq$20\% when averaged over the filament region, implying the presence of ordered magnetic fields. The high polarization and the intensity and spectral shape of the emission, they claim, are the first pieces of observational evidence of particle acceleration around the cosmic filaments. This synchrotron cosmic web from strong structure-formation shocks may have, in fact, been previously observed in the diffuse radio synchrotron background, as we discuss in Section~\ref{sec:RSB}.

While radio observations have provided evidence of strong shocks and amplified magnetic fields, the lack of obvious sources of secondary gamma rays and neutrinos constrain some UHECR source models. The negative source evolution expected for accretion shocks contrasts sharply with the strong positive evolution typically associated with compact sources, offering a robust criterion for differentiation through studies of the cosmic ray spectrum, composition, and anisotropy. Models of UHECR acceleration that involve the rates of some types of AGNs, gamma-ray bursts, or star formation have a source evolution proportional to $(1+z)^m$ for $m \geq 0$ for $z \leq 1$. In contrast, accretion shocks have a negative source evolution due to the long timescales involved for cosmic rays above 100\,EeV. In fact, the gamma-ray background and neutrino background exclude a strong positive source evolution \citep{2017ApJ...839L..22G,2023JCAP...05..024A}.

Moreover, if accretion shocks are the dominant source of UHECRs, they should produce a distinct spectrum of cosmogenic neutrinos compared to models in which UHECRs originate from compact sources, which exhibit positive cosmic evolution and more efficient early-Universe neutrino production. Unlike astrophysical neutrinos, which are produced at the source via interactions within the acceleration region, cosmogenic neutrinos originate from UHECR interactions with the CMB or extragalactic background light during propagation. However, if UHECRs experience prolonged acceleration in accretion shocks, the distinction between astrophysical and cosmogenic neutrinos becomes less clear, as interactions could occur both during acceleration and after escape. The long accumulation time and negative source evolution of accretion shocks imply a cosmogenic neutrino spectrum that peaks at lower redshifts compared to models with strong positive evolution. This prediction may be tested with existing and upcoming neutrino observatories such as IceCube, KM3NeT, RNO-G, P-ONE, and GRAND. A possible indication of EeV neutrinos has emerged with the recent detection of a muon event, KM3-230213A, by KM3NeT. This event, interpreted as a muon neutrino with an energy in the range of 72\,PeV to 2.6\,EeV, has a median energy of 220\,PeV based on detector simulations \citep{2025Natur.638..376K}. If confirmed, this detection could provide insight into the spectrum and evolution of UHECR sources, although its inferred flux appears to be high relative to most cosmogenic neutrino models and is in slight tension with the absence of a detection from IceCube \citep{2025arXiv250201963I,2025arXiv250204508L}.

\subsection{Synchrotron emission from downstream electrons}
\label{sec:RSB}

The primary electrons accelerated at the shocks provide another powerful diagnostic with the integrated emission of the entire cosmic web. The electron synchrotron emission is one of the best ways to observationally discriminate an accretion-shock origin from an AGN origin of UHECRs because the angular distributions of the two sources are very different \citep{Simeon:20233q}.

The diffuse and hard radio synchrotron emission from electrons accelerated at accretion shocks may partially explain the observation of a temperature excess in the radio synchrotron background (also known as the cosmic radio background) from 22~MHz to 10~GHz, implying a spectral flux density $S_{\nu} \propto \nu^{-\alpha}$ with an index of $\alpha \simeq 0.6$ \citep{2011ApJ...734....5F, 2011ApJ...734....6S}. This spectrum, implying an electron power-law index of $p = 2.2$, is harder than those of commonly observed radio halos and relics, and it is difficult to explain with known sources \citep{2010MNRAS.409.1172S,2023PASP..135c6001S}. More recent studies have not found discrete sources and suggest the explanation is either foreground confusion or an unknown diffuse component \citep{2023MNRAS.521..332T}.
Because synchrotron emissivity scales with $B^2$, the electrons in the thin, high-field regions immediately downstream of the accretion shocks dominate the observed flux, yielding a naturally hard spectral shape without electron cooling. The extremely low density and extended nature are the most prominent features that make them so hard to detect.

We now compare the power of the radio synchrotron background with the cosmological average power of the shocks in our simulation. The energy density of the radio background per natural logarithmic frequency interval, evaluated at $\nu_{\rm r}$, is
\begin{eqnarray}
\nu_{\rm r} U_{\nu_{\rm r}} = 1.17\, \frac{8\pi k_{\rm B} \nu_{*}^{3}}{c^3} \left( \frac{\nu_{\rm r}}{\nu_{*}} \right)^{0.4},
\end{eqnarray}
where $\nu_* = 1\,{\rm GHz}$ \citep{2010MNRAS.409.1172S}. When integrated up to 10~GHz, this is an energy density of $\rho_{\rm RSB} \approx 9 \times 10^{-20} \,{\rm J\,m}^{-3}$. They relate the radio synchrotron energy density to the emissivity $j_{\nu_{\rm r}}$ of the relativistic electrons as follows:
\begin{eqnarray}
[\nu_{\rm r} U_{\nu_{\rm r}}] = [\nu_{\rm r} j_{\nu_{\rm r}}] \frac{4\pi}{H_0} \int \frac{ F_{\rm syn}(z) dz}{(1+z)^{(p+1)/2} E(z)},
\end{eqnarray}
where $F_{\rm syn}(z)$ describes the evolution of the product $U_B \times k_{\rm e}$ and $E(z) \equiv \sqrt{\Omega_{\rm M} (1+z)^3 + \Omega_{\Lambda}}$ for a spatially flat cosmology. We set $F_{\rm syn}(z) = 1$, assuming that this factor is of order unity. We use the same cosmological parameters as in our simulation. We integrate up to $z = 1$, but integrating to higher $z$ hardly affects the answer. We solve for $[\nu_{\rm r} j_{\nu_{\rm r}}]$ and integrate over the same frequency range as above to get the power density of the radio synchrotron background to be $4.2 \times 10^{-38}\, {\rm W\,m}^{-3}$, or 0.12\% of $\mathcal{P}_{\rm LSSS}$, the cosmological average power density of large-scale shocks at $z = 0$ as given in Section~\ref{sec:sims}. Using the electron-to-proton energy ratio of $\kappa_{\rm ep} \approx 0.05$ for $q_{\rm e} = q_{\rm p} = 2.2$ \citep{2014A&A...567A.101P} and a cosmic-ray efficiency of $\eta_{\rm CR} \approx 0.1$, the power density put into relativistic electrons is ${\cal P}_{\rm e} \approx \kappa_{\rm ep}\eta_{\rm CR}{\cal P}_{\rm LSSS} \approx 4 \, {\cal L}_{\rm RSB}$. Although there is certainly some evolution to the power density of large-scale shocks, we expect the evolution to be relatively flat up to $z = 1$.

Furthermore, if strong accretion shocks are producing the cosmic radio background, the magnetic field in the region producing that synchrotron emission should be greater than 1~$\mu$G. \citet{2010MNRAS.409.1172S} found that the nonthermal electrons would inverse Compton scatter the CMB field and overproduce X-rays and $\gamma$-rays if the magnetic field is any less than about 1~$\mu$G. Though not stated in that paper, that limit likely comes from integrating the resultant inverse Compton emission up to redshifts higher than expected if the source were strong accretion shocks, which would not be bright sources at $z \gtrsim 3$. Therefore, this constraint may be weaker, allowing for the slightly weaker magnetic fields in agreement with observations mentioned in Section\,\ref{subsec:obs_constraints}. Throughout this paper, we have been assuming a total amplified magnetic field of 1~$\mu$G as a reference value based on the confinement and acceleration requirements, but this value is also flexible if $u_1 > 1000\,{\rm km\,s}^{-1}$. By linking the magnetic field near the shocks to the observational signatures of the accompanying electrons, we have an independent check on the minimum value of the magnetic field consistent with strong acceleration of UHECRs. More work is needed to determine whether the angular power spectrum of synchrotron emission from cluster shocks matches the measured angular power spectrum of the radio synchrotron background \citep{2022MNRAS.509..114O,2023MNRAS.523.5034C}.

While the energetics show that accretion shocks could, in principle, account for the observed radio synchrotron background, the observed radio morphologies indicate that their contribution is much fainter than that of more common radio relics and halos. The Coma cluster is the only system with a radio feature tentatively associated with an accretion shock \citep{2022ApJ...933..218B}, suggesting that such emission is exceptionally faint and rare, and thus unlikely to make a significant contribution to the diffuse radio background.

\section{Discussion}
\label{sec:discussion}

The primary motivation for this model is the continuous and relatively smooth cosmic ray spectrum over a vast energy range. Aside from minor breaks in the spectral index, no clear transition separates Galactic from extragalactic cosmic rays. One possible explanation for this continuous spectrum is that the same process---DSA at strong, nonrelativistic, collisionless shocks---reaccelerates cosmic rays at multiple sites, allowing them to reach progressively higher energies. We do not attempt to model the entire spectrum, as doing so would require modeling the escape from each shock type and the propagation from the source galaxies to our Galaxy.

Such modeling of the full spectrum would also track the composition, which increases in mean mass between about 1\,PeV and $\sim$400\,PeV. Galactic cosmic ray sources presumably have a maximum rigidity of 3 to 5\,PV, allowing them to accelerate iron up to $\sim$100\,PeV. Consistent with this picture, the data show that protons are less abundant than helium and carbon from 1\,PeV to at least 20\,PeV, and an iron knee appears at 80\,PeV \citep{2013APh....47...54A}. Near the so-called second knee around 400\,PeV, as much as 40\% of the cosmic rays could be from elements as massive as lead \citep{2013FrPhy...8..748G,2020PhR...872....1B}.

Yet, protons are still present in this region of heavy composition where typical Galactic sources cannot accelerate them. Either atypical Galactic sources accelerate protons without heavy elements, or, as we argue, those protons come from nearby accretion shocks surrounding the filament or sheet that hosts the Local Group. These shocks reaccelerate the mixed composition of galactic cosmic rays and produce a steady downstream flux that is mostly protons below the maximum rigidity $\sim$1\,EV for filament shocks.

From 100\,PeV to $\sim$4\,EeV, the composition becomes increasingly light because the heavier species are less abundant in the Galactic cosmic ray distribution, while protons are still abundant at extragalactic sources. Above the maximum energy of protons at filament accretion shocks ($\sim$1\,EeV), one might expect the mean mass to increase again, similar to how Galactic cosmic rays became heavier at lower energies. However, the heavy downstream flux from nearby shocks is exceeded by the hard spectrum escaping upstream from more distant accretion shocks. The light-to-heavy cycle restarts before significant numbers of heavy cosmic rays from nearby filament shocks appear.

Above the ankle ($\sim$4\,EeV), the composition becomes increasingly heavy \citep{2017JCAP...04..038A,2023JCAP...05..024A}. If nuclei as heavy as lead become prominent around 400\,PeV, as previously mentioned, then they should also be present at the highest-energies ($>$200\,EeV) at the source. Detecting these extremely rare heavy nuclei---those that survive long acceleration and propagation times---will require much larger collecting areas in future experiments.

In addition to the energy spectrum and composition, the low levels of anisotropy are consistent with our model. The amplitude of the dipole anisotropy remains below 0.01 for energies below 10\,PeV and points toward the Galactic Center between 100\,TeV and 10\,PeV \citep{2017PrPNP..94..184A}. Above 10\,PeV, the dipole direction shifts because more cosmic rays arrive from outside the Galaxy. The energy dependence of the dipole amplitude mirrors the energy scaling of the diffusion tensor, suggesting that 1\,EeV cosmic rays experience a level of diffusion in the extragalactic magnetic fields similar to that of 10\,PeV cosmic rays in the Galactic magnetic field. Intergalactic magnetic fields around the Local Group were likely seeded by early galactic outflows and could reach $\sim$10\,nG. In this case, the relevant scattering scale, set by the gyroradius for Bohm diffusion, would be $\sim$100\,kpc, smaller than the typical distance to the nearest accretion shocks. Thus, the observed amplitude and direction of anisotropy indicate that the transition from Galactic to extragalactic cosmic rays aligns with the transition inferred from the composition. Better measurements of the anisotropy in this energy range would further improve our understanding of this transition region.

A hierarchical shock model can explain most of the cosmic ray observables and their secondaries using known mechanisms. The somewhat contentious requirement of 1\,$\mu$G magnetic fields upstream of accretion shocks is needed only to explain cosmic rays above a few EeV. A natural place to invoke a different type of cosmic ray source is the hard spectrum above the ankle ($\sim$4\,EeV), the most dramatic change in the spectrum aside from solar modulation below 1\,GeV.

Other potential UHECR sources---AGN jets and gamma-ray bursts---are certainly able to accelerate particles to high energies, but their extreme luminosities also present challenges. As some of the brightest sources in the Universe, their intense photon backgrounds make it difficult for heavy nuclei to survive. In addition to the intense radiation fields, UHECRs from these sources must escape environments with higher magnetic fields and denser gas before reaching intergalactic space, whereas accretion shocks accelerate particles directly at the boundaries with voids.

The temperature downstream of accretion shocks often exceeds $10^6$\,K, but the extremely low density and optical depth of the shocked gas ensure that photodisintegration is negligible. In contrast, the other luminous UHECR candidates are better suited for producing astrophysical neutrinos, which require a source with a high proton opacity. Given the different constraints on UHECR and neutrino production, escape, and observed spectra, a single source class is unlikely to simultaneously account for both UHECRs and astrophysical neutrinos.

Strong accretion shocks must exist and accelerate particles to high energies. The key uncertainties are the maximum energy they can achieve and the spectral index of the particles that reach Earth. To better constrain these factors, we are investigating the escape upstream of curved shocks \citep{2023arXiv230909116B}.

\section{Summary and Conclusions}
\label{sec:conclusion}

We have presented a hierarchical model of DSA that builds upon many existing ideas in the literature to explain the full cosmic ray spectrum and the secondary signatures that accompany it.

This work incorporates both observations of the local universe and a cosmological simulation, but some conclusions of these two lines of research are in tension with each other. Compared to simulations, which typically adopt a lower $H_0$ based on CMB observations, local measurements suggest more massive clusters, faster flow speeds, and a higher $H_0$. If local observations are indeed more accurate, they would favor accretion shocks as viable sources of UHECRs.

In summary, we highlight the key conclusions of our paper.
\begin{enumerate}
    \item We propose a hierarchical shock acceleration model---from supernova remnants and galactic wind shocks to large-scale accretion shocks around filaments and clusters---that accounts for the observed trends in the cosmic ray spectrum and composition from $\sim$1\,GeV to $\sim$200\,EeV. 
    
    \item The highest-energy UHECRs are predominantly heavy nuclei (iron group) with rigidities of 2–9\,EV. The observed mixed composition arises from the sequential reacceleration of galactic cosmic rays at increasingly larger shocks.
    
    \item Achieving these rigidities requires microgauss-level magnetic fields upstream of the shocks. Several mechanisms---early-Universe galactic outflows, cosmic ray streaming instabilities, and a small-scale turbulent dynamo---plausibly act in concert to amplify magnetic fields to $\sim$1\,$\mu$G, particularly in localized regions where the most efficient particle acceleration occurs.

    \item Our high-resolution radiation hydrodynamic AMR simulation of a $256\,{\rm Mpc}\,h^{-1}$ co-moving volume reveals that the high-Mach, low-density accretion shocks---expected at the outskirts of clusters and filaments---yield a volume-averaged power density of $\mathcal{P}_{\rm LSSS} \sim 10^{40} \,{\rm erg\,s}^{-1}\,{\rm Mpc}^{-3}$, well above the UHECR luminosity requirement. We find a bimodal distribution of shock speeds (clusters and filaments) with typical upstream speeds for clusters around $1000\,{\rm km\,s}^{-1}$ and rare regions exceeding $5000\,{\rm km\,s}^{-1}$.
    
    \item Filament shocks, which have more shock surface area per cosmological volume and which are likely to be at most a few megaparsecs from our Galaxy, contribute significantly to the cosmic ray flux in the intermediate rigidity range between the Galactic petavolt sources and the circumcluster exavolt sources. The downstream flux from these shocks is naturally softer than the upstream flux escaping from cluster shocks because higher-rigidity particles preferentially escape from the filamentary environment into surrounding voids and because variations in shock conditions produce a spread in rigidity cutoffs. This spectral dichotomy between upstream and downstream escape explains the transition from a soft, light composition to a hard, heavy composition at the highest energies.

    \item Accretion shocks are a counter-evolutionary source, exhibiting lower UHECR luminosity at high redshifts. This feature naturally avoids conflicts with constraints from the gamma-ray and neutrino backgrounds.

    \item Synchrotron and inverse Compton emission from primary and secondary electrons remain consistent with constraints on both diffuse emission and individual clusters. In fact, compelling evidence is accumulating that these accretion shocks are seen in radio and gamma rays.

    \item Electrons accelerated at these shocks emit synchrotron radiation, which may contribute to the radio synchrotron background between 22\,MHz and 10\,GHz. More work is needed to compare the anisotropy of the synchrotron background to the expected anisotropy from accretion shocks. Future observations with DSA\,2000 and SKA may directly detect these accretion shocks.
\end{enumerate}

\section*{Acknowledgments}
The authors thank D. Allard, T. Porter, S. Allen, N. Werner, A. Simionescu, O. Urban, J. Singal, D. Caprioli, F. Aharonian, S. Martin-Alvarez, and P. Helbig for helpful discussions and feedback. K.S.S.B. was supported by XSEDE computing grants TG-AST190001 and TG-AST180052, the Stampede2 supercomputer at the Texas Advanced Computing Center, a Porat Postdoctoral Fellowship at Stanford University, and a NASA Hubble Postdoctoral Fellowship at Harvard University. We acknowledge support from the Simons Foundation (MP-SCMPS-00001470, N.G., R.B.). We thank the anonymous referee for thorough and constructive comments.

\software{Enzo \citep{2014ApJS..211...19B},
    Music \citep{2011MNRAS.415.2101H},
    Rockstar \citep{2013ApJ...762..109B},
    Consistent Trees \citep{2013ApJ...763...18B}.
    }


\section*{ORCID iDs}

\noindent Paul Simeon \orcidlink{0000-0001-7763-4405} \href{https://orcid.org/0000-0001-7763-4405}{https://orcid.org/0000-0001-7763-4405} \\
Noémie Globus \orcidlink{0000-0001-9011-0737} \href{https://orcid.org/0000-0001-9011-0737}{https://orcid.org/0000-0001-9011-0737} \\
Kirk S. S. Barrow \orcidlink{0000-0002-8638-1697} \href{https://orcid.org/0000-0002-8638-1697}{https://orcid.org/0000-0002-8638-1697} \\
Roger Blandford \orcidlink{0000-0002-1854-5506} \href{https://orcid.org/0000-0002-1854-5506}{https://orcid.org/0000-0002-1854-5506}

\bibliographystyle{aasjournal}


\end{document}